\documentclass[a4paper,11pt]{article}
\pdfoutput=1
\usepackage{jheppub}
\usepackage{color,bm,comment,braket}
\usepackage[T1]{fontenc}
\usepackage{comment}
\newcommand{\del}{\partial}
\newcommand{\beq}{\begin{eqnarray}}
\newcommand{\eeq}{\end{eqnarray}}

\newcommand{\tr}{\mathop{\mathrm{tr}}}
\newcommand{\SU}{\text{SU}}
\newcommand{\U}{\text{U}}
\newcommand{\rmi}{\text{i}}
\newcommand{\rme}{\text{e}}
\newcommand{\rmd}{\text{d}}

\begin{document}

\preprint{YGHP-23-06}

\title{
 Non-Abelian chiral soliton lattice in rotating QCD matter: 
Nambu-Goldstone and excited modes 
}

\author[a,b]{Minoru Eto}
\affiliation[a]{%
Department of Physics, Yamagata University, 
Kojirakawa-machi 1-4-12, Yamagata,
Yamagata 990-8560, Japan}
\affiliation[b]{Research and Education Center for Natural Sciences, Keio University, 4-1-1 Hiyoshi, Yokohama, Kanagawa 223-8521, Japan}
\emailAdd{meto@sci.kj.yamagata-u.ac.jp}

\author[c,b]{Kentaro Nishimura}
\affiliation[c]{International Institute for Sustainability with Knotted Chiral Meta Matter(SKCM$^2$), Hiroshima University, 1-3-2 Kagamiyama, Higashi-Hiroshima, Hiroshima 739-8511, Japan}
\emailAdd{nishiken@hiroshima-u.ac.jp}

\author[d,b,c]{and Muneto Nitta
}
\emailAdd{nitta@phys-h.keio.ac.jp}
\affiliation[d]{Department of Physics, Keio University, 4-1-1 Hiyoshi, Kanagawa 223-8521, Japan}

\abstract{
The ground state of QCD with two flavors
at a finite baryon chemical potential 
under rapid rotation
is a chiral soliton lattice (CSL) 
of the $\eta$ meson,  
consisting of a stack of 
sine-Gordon solitons 
carrying a baryon number, 
due to 
the anomalous coupling of the $\eta$ meson 
to the rotation.
In a large parameter region, 
the ground state becomes a non-Abelian CSL, 
in which 
due to the neutral pion condensation 
each $\eta$ soliton decays into a pair of 
non-Abelian sine-Gordon solitons 
carrying $S^2$ moduli originated from 
Nambu-Goldstone (NG) modes localized around it,
corresponding to the spontaneously broken 
vector symmetry SU$(2)_{\rm V}$.
There, the $S^2$ modes of neighboring solitons are anti-aligned,
and these modes should propagate in the transverse direction of the lattice due to the interaction between the $S^2$ modes of neighboring solitons.
In this paper, 
we calculate excitations 
including gapless NG modes and excited modes 
around non-Abelian and Abelian ($\eta$) CSLs, 
 and find 
 three gapless NG modes 
 with linear dispersion relations (type-A NG modes):
 two isospinons ($S^2$ modes) 
 and a phonon 
 corresponding to 
 the spontaneously broken vector  SU$(2)_{\rm V}$ and translational symmetries around 
 the non-Abelian CSL, respectively, and only a phonon for 
 the Abelian CSL because of the recovering SU$(2)_{\rm V}$.
 We also find in the deconfined phase that the dispersion relation of the isospinons becomes of the Dirac type, {\it i.~e.~} linear even at large momentum.}

\maketitle

\section{Introduction}

The strong interaction, one of the fundamental forces, can be described by quantum chromodynamics (QCD), 
the theory for quarks and gluons. 
In the vacuum, one does not observe quarks, 
but there are only hadrons, {\it i.~e.} 
baryons made of three quark-bound states 
and mesons made of quark--anti-quark bound states.
In fact, 
a lattice formulation of QCD can show such quark confinement 
at the quantitative level.
However, because of the sign problem, 
the lattice QCD cannot describe 
a situation at finite baryon density relevant for neutron star interiors 
and heavy-ion colliders. 
In order to explore such a situation, 
much attention has been paid to 
QCD phase diagram at finite baryon density 
under external fields such as 
strong magnetic field and rapid rotation \cite{Fukushima:2010bq}.

Low-energy dynamics of QCD can be described in terms of Nambu-Goldstone(NG) bosons, 
that is pions, when the chiral symmetry is spontaneously broken.
Theoretically, 
it is a chiral Lagrangian or the chiral perturbation theory (ChPT), which is unique up to some constants,
the pion's decay constant, 
quark masses, and so on \cite{Scherer:2012xha,Bogner:2009bt}.  
In the presence of an external magnetic field,
the chiral Lagrangian is accompanied by the anomalous coupling of the neutral pion $\pi_0$ to the magnetic field via the chiral anomaly 
\cite{Son:2004tq,Son:2007ny}
through  
the Goldstone-Wilczek current \cite{Goldstone:1981kk,Witten:1983tw}.
Then, 
the ground state of QCD with two flavors (up and down quarks) 
at a finite baryon chemical potential 
in the presence of a sufficiently strong magnetic field 
was found to be 
a chiral soliton lattice (CSL) 
consisting of a stack of 
domain walls or sine-Gordon solitons 
carrying a baryon number 
\cite{Son:2007ny,Eto:2012qd,Brauner:2016pko}.\footnote{
It is known that some inhomogeneous states modulated along one direction are unstable under transverse fluctuations 
assuming the SO(3) rotational symmetry of the system, 
which is called the Landau-Peierls instability 
\cite{Lee:2015bva,Hidaka:2015xza}.
However, in this case, such an instability is absent 
because the SO(3) rotational symmetry is explicitly broken 
by magnetic field or rotation.
}
However, such a CSL is unstable against 
a charged pion condensation 
in a region of higher density and/or stronger 
magnetic field \cite{Brauner:2017uiu}.
One of the possibilities for the fate of the instability is an Abrikosov's vortex lattice or baryon crystal \cite{Evans:2022hwr,Evans:2023hms}. 
On the other hand, we showed 
in refs.~\cite{Eto:2023lyo,Eto:2023wul}
that there appears to be a lower energy state before going to such an unstable region, 
{\it i.~e.}~ the domain-wall Skyrmion phase
in which Skyrmions are created on top of the solitons 
in the ground state.
The instability of the CSL is present 
in the region where it is metastable and  
the ground state is the domain-wall Skyrmion phase \cite{Eto:2023wul}.
Possible relations between Skyrmion crystals 
in the absence of the magnetic field 
and the CSL phase were discussed in 
refs.~\cite{Kawaguchi:2018fpi,Chen:2021vou,Chen:2023jbq}. 
The other topics of the CSLs include, for instance 
CSLs  
under thermal fluctuation \cite{Brauner:2017uiu,Brauner:2017mui,Brauner:2021sci,Brauner:2023ort},  
quantum nucleation of CSLs 
\cite{Eto:2022lhu,Higaki:2022gnw},
and quasicrystals \cite{Qiu:2023guy}.

\begin{figure}[tb]
    \centering
    \includegraphics[width=15.0cm]{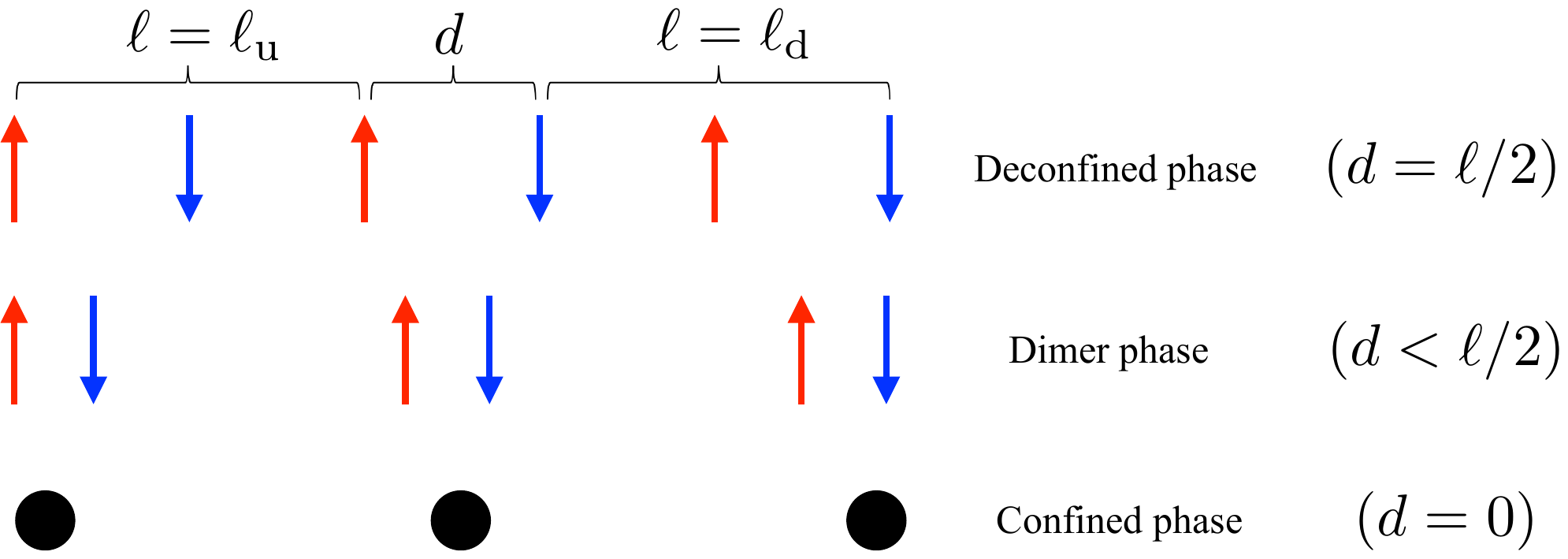}
    \caption{
    Schematic picture of an arrangement of the up (red arrow) and down (blue arrow) solitons.
    Since $\ell_+=\ell_-$, the state where the up and down solitons farthest apart satisfies $d=\ell/2$.
    This state is called the deconfined phase.
    On the other hand, the state where the up and down solitons are completely overlapped (the black circle in this picture) is called the confined phase.
    The intermediate state satisfies $d<\ell/2$.
    }
    \label{fig:phases}
\end{figure}

Here, we concentrate on another 
important axis, rotation, in the phase diagram of QCD.
Quark-gluon plasmas produced in non-central heavy-ion collision experiments at the Relativistic Heavy Ion Collider (RHIC) reach the largest vorticity observed thus far, of the order of $10^{22}/{\rm s}$ \cite{STAR:2017ckg,STAR:2018gyt}. 
This has triggered significant attention to rotating QCD matter in recent years 
\cite{Chen:2015hfc,Ebihara:2016fwa,Jiang:2016wvv,Chernodub:2016kxh,Chernodub:2017ref,Liu:2017zhl,Zhang:2018ome,Wang:2018zrn,Chen:2019tcp,Chernodub:2020qah, Chernodub:2022veq, Huang:2017pqe,Nishimura:2020odq,Chen:2021aiq,
Eto:2021gyy}. 
In parallel to the case of a strong magnetic field, 
the CSL phase in rotating QCD matter was  studied in refs.~\cite{Huang:2017pqe,Nishimura:2020odq,Eto:2021gyy,Chen:2021aiq,Eto:2023tuu}: 
a contribution of the chiral anomaly to the $\eta'$ meson was obtained by matching with 
the chiral vortcical effect (CVE) 
\cite{Vilenkin:1979ui,Vilenkin:1980zv,Son:2009tf,Landsteiner:2011cp,Landsteiner:2012kd,Landsteiner:2016led}
in terms of mesons \cite{Huang:2017pqe,Nishimura:2020odq}.
Due to the anomalous term, 
the ground state in a certain parameter region 
was found to be 
another type of CSL made of the $\eta'$ meson 
\cite{Huang:2017pqe,Nishimura:2020odq} 
instead of that of the neutral pion $\pi_0$ 
in the magnetic field.
While in ref.~\cite{Nishimura:2020odq} the pion fields 
are set to be zero, 
it cannot be justified always.
In fact, it was found in ref.~\cite{Eto:2021gyy} 
that in a large parameter region 
in the two-flavor case\footnote{
In the two-flavor case,
the $\eta$ meson plays a role in constituting a CSL instead of the $\eta'$ meson in the three-flavor case.
}
the neutral pion condensation occurs 
in the CSL  
to turn it a {\it non-Abelian CSL}, 
where a single $\eta$ soliton 
decays into a pair of 
non-Abelian sine-Gordon solitons. 
For a single non-Abelian soliton, 
the $\eta$ meson changes from $0$ to $\pi$ 
and the neutral pion $\pi_0$ changes from 
$0$ to $\pi$ 
so that the U$(2)$ group element is single-valued in the periodic boundaries,  
in contrast to 
a single $\eta$ soliton
for which only the $\eta$ meson 
changes from $0$ to $2\pi$ while 
the pions are zero. 
Because of a nontrivial $\pi_0$ profile, 
the vector symmetry SU$(2)_{\rm V}$ 
unbroken in the vacuum
is further broken spontaneously into 
a U$(1)$ subgroup in the vicinity of the soliton, resulting in 
${\mathbb C}P^1 [\simeq {\rm SU}(2)/{\rm U}(1)]$ 
NG modes localized around each soliton, 
thus dubbed a non-Abelian sine-Gordon soliton \cite{Nitta:2014rxa,Eto:2015uqa} 
(see also refs.~\cite{Nitta:2015mma,Nitta:2015mxa,Nitta:2022ahj} for futher study). 
The ${\mathbb C}P^1$ modes at two neighboring solitons repel each other, and thus they are antialigned. 
Thus, if one calls one soliton an up-soliton, 
then its neighboring solitons 
can be called down-solitons, 
and the up and down solitons appear alternately in the whole lattice.
Thus, non-Abelian CSL can be regarded as 
a Heisenberg anti-ferromagnetic spin chain. 
Such a non-Abelian CSL can be distinguished by 
a ${\mathbb Z}_2$ symmetry into the two classes: the deconfined and dimer phases.
In the deconfined phase, 
an up-soliton and down-soliton repel each other 
in real space, 
and thus up and down solitons appear alternately 
with the same distances, 
as illustrated in 
the upper line in fig.~\ref{fig:phases}.
In the dimmer phase, 
they attract each other at large distances 
and repel at short distances, 
and thus constitute a molecule 
as illustrated in 
the middle line in fig.~\ref{fig:phases}. 
In the confined phase, 
the up-soliton and down-soliton 
attract in the whole separation,  
and they completely overlap 
to constitute an $\eta$ soliton, 
as illustrated in 
the bottom line in fig.~\ref{fig:phases}.
This is nothing but the $\eta$ CSL, which is also 
called an Abelian CSL. 
These three phases appear in 
the phase diagram spanned by 
the ratio between the 
decay constants of the $\eta$ meson and pions 
and the ratio between the anomaly coefficient 
and pion mass 
 \cite{Eto:2021gyy}.
Furthermore, as in the case of the strong magnetic field,
a domain-wall Skyrmion phase appears in a larger chemical potential region inside the non-Abelian CSL \cite{Eto:2023tuu}.

In this paper, we study excitations 
including gapless 
NG modes and massive modes 
around 
the three phases of CSLs mentioned above, 
the confined, deconfined and dimer phases. 
We also consider a noninteractive case, 
for which up and down solitons do not interact, 
and consequently, up solitons form a lattice, 
and down solitons form another lattice 
independently.
In the classification of NG modes, 
type-A(B) NG modes correspond 
to the same number (a half) of broken generators, 
typically having linear (quadratic) dispersion relations 
\cite{
Watanabe:2012hr,
Hidaka:2012ym,
Watanabe:2014fva,
Takahashi:2014vua}, 
see refs.~\cite{
Watanabe:2019xul,
Beekman:2019pmi} for a review. 
For example, magnons in (anti-)ferromagnets 
are type-B(A) NG modes with 
quadratic (linear) dispersion relations. 
Another example of NG modes in a soliton lattice is a phonon-like excitation propagating in transverse directions of a vortex lattice, called a Tkachenko mode \cite{Tkachenko1,Tkachenko2,Tkachenko3,Du:2022xys}, 
which is a type-B NG mode 
\cite{Watanabe:2013iia}.
In the case of a non-Abelian CSL, 
 non-Abelian solitons form a lattice,  
 in which the ${\mathbb C}P^1$ modes of 
a pair of neighboring solitons are anti-aligned, and 
 the ${\mathbb C}P^1$ modes 
 as well as a phonon should be able to 
 propagate in the transverse direction of the lattice. 
 Therefore, we work out fluctuations 
 around the non-Abelian and Abelian CSLs 
 and find 
 three gapless NG modes 
 with linear dispersion relations (type-A):
 two isospinons (the ${\mathbb C}P^1$ modes) 
 corresponding to 
 the spontaneously broken SU$(2)_{\rm V}$ symmetry, 
 beside a phonon corresponding to
  the spontaneously broken 
  translational symmetry.
In particular, the dispersion relation of the isospinons is of the Dirac type, {\it i.~e.~} linear even at large momentum 
in the deconfined phase, 
and is not so 
in the dimer phase. 
 On the other hand, in the confined phase,
 we find only one gapless NG mode (a phonon) 
 as expected,
 since the vecotr symmetry SU$(2)_{\rm V}$ is not broken. 
 
This paper is organized as follows.
In sec.~\ref{sec:NACSL} we summarize 
non-Abelian CSLs.
In ssc.~\ref{sec:dispersion}, 
we study fluctuations around the CSLs 
to obtain dispersion relations of 
NG modes and excited modes.
Sec.~\ref{sec:summary} is devoted to 
a summary and discussion.

\section{Non-Abelian chiral soliton lattices} \label{sec:NACSL}
In this section, we review the non-Abelian CSL in two-flavor QCD at finite baryon chemical potential $\mu_\textrm{B}$ and angular velocity $\bm{\Omega}=\Omega \hat{\bm{z}}$ with flavor symmetric masses, $M=\textrm{diag}(m,m)$.

\subsection{General setup}
The chiral symmetry U$(2)_{\rm L} \times $U$(2)_{\rm R}$ of the two flavor QCD is
\begin{gather}
q_\textrm{R} \to \rme^{-\rmi \tau_0\theta_0^\textrm{R}} V_\textrm{R}q_\textrm{R} \,, \qquad
q_\textrm{L} \to \rme^{-\rmi \tau_0\theta_0^\textrm{L}} V_\textrm{L}q_\textrm{L} \,,
\end{gather}
where $V_\textrm{R,L} \equiv \exp (-\rmi \tau_A\theta^{\textrm{R}, \textrm{L}}_A)$ are $\SU(2)$ matrices with 
 $\U(2)$ generators $\tau_a$ with the normalization $\tr(\tau_a\tau_b)=2\delta _{ab}$.
Here and below, the capital Latin letters run over $1, 2, 3$, and the small letters run over $0, 1, 2, 3$.
We assume that this chiral symmetry is spontaneously broken down to the vector symmetry $\U(2)_\textrm{V}$.
Consequently, the $\eta$ meson and pions appear as  NG bosons which can be parametrized as
\begin{gather}
U = \Sigma \exp \left(\frac{\rmi \tau_0 \eta}{f_{\eta}} \right) \,, \qquad
\Sigma = \exp \left(\frac{\rmi \tau_A\pi_A}{f_{\pi}} \right) \in \SU(2) \,,
\end{gather}
with the decay constants $f_{\eta, \pi}$ of the $\eta$ and $\pi$ mesons.
The field $\Sigma$ transforms under $\SU(2)_\textrm{L} \times \SU(2)_\textrm{R}$ as
\begin{gather}
\Sigma \to V_\textrm{L} \Sigma V_\textrm{R}^{\dag} \,,
\label{eq:transf_Sigma}
\end{gather}
while $\eta$ transforms under $\U(1)_\textrm{A}$ as
\begin{gather}
\eta \to \eta + 2f_{\eta} \theta_0
\end{gather}
where $\theta_0 \equiv \theta_0^\textrm{R} = -\theta_0^\textrm{L}$.

The effective Lagrangian for 
the $\eta$ and $\pi$ mesons is given by
\begin{align}
    \mathcal{L}
    &= \frac{1}{4}f^2_{\pi} \tr(\del_{\mu}\Sigma \cdot \del^{\mu}\Sigma^{\dag})
    + \frac{1}{2}(\del_{\mu}\eta)^2 \notag \\
    &+ \frac{Bm}{2} \tr(U+U^{\dag}-2\bm{1}_2)
    + \frac{A}{2} (\det U + \det U^{\dag}-2\bm{1}_2)
    + \frac{\Omega \mu_{\textrm{B}}}{2\pi^2N_{\textrm{c}}}\del_z\left(\frac{\eta}{f_{\eta}} \right)
    \label{lagrangian} \,.
\end{align}
More precisely, for the speeds of propagation of $\pi$ and $\eta$  in a medium are different in general, we should replace $\del_\mu\del^\mu \to \del_t^2 - v_{\pi,\eta}^2 \nabla^2$ in the above Lagrangian. However, we will assume $v_\pi = v_\eta = 1$ in this paper for simplicity as in Ref.~\cite{Brauner:2016pko}.
The first and second terms are the kinetic terms of the $\pi$ and $\eta$ mesons, respectively. 
They are invariant under ${\rm S[U(2)_L\times U(2)_R]}$.
The third term originates from explicit symmetry breaking due to the quark masses $m$. 
The pion mass $m_{\pi}$ is related 
to the quark mass by
$Bm= f_\pi^2 m_\pi^2/4$.
It is invariant under the vector-like symmetry ${\rm SU(2)_V}$ (taking $V_L=V_R$ in eq.~(\ref{eq:transf_Sigma})).
The forth term which is invariant under ${\rm SU(2)_L\times SU(2)_R}$ corresponds to the $\U(1)_{\textrm{A}}$ anomaly.
The last term represents the anomalous coupling of the $\eta$ meson to the rotation via the chiral anomaly.
This term is derived by the anomaly matching for the CVE in terms of mesons \cite{Huang:2017pqe,Nishimura:2020odq}.
Hence, the Lagrangian symmetry is generally ${\rm SU(2)_V}$.

The target space ${\rm U(2)}$ is not simply connected. Since the first homotopy group is non-trivial as $\pi_1[{\rm U(2)}] = \mathbb{Z}$, this system possesses topological solitons, 
non-Abelian sine-Gordon solitons 
\cite{Nitta:2014rxa,Eto:2015uqa}. 
We will emphasize difference between U$(1) \times $SU$(2)$ and U$(2) \simeq \frac{{\rm U}(1)\times {\rm SU}(2)}{\mathbb{Z}_2}$.

For convenience, we will use the following quantities,
\begin{gather}
\textrm{sin}\beta \equiv \frac{A}{C} \,, \qquad \textrm{cos}\beta \equiv \frac{2Bm}{C} \,, \qquad C \equiv \sqrt{A^2 + (2Bm)^2} \,.
\end{gather}
We assume both $A$ and $B$ are positive, so the range of $\beta$ is $0 \le \beta \le \pi/2$.
The potential term can be rewritten as  
\begin{gather}
    \mathcal{L}_\textrm{pot}
    \equiv -V_{\textrm{pot}}
    = C\left[
    \frac{1}{4}\textrm{cos}\beta \cdot \tr(U+U^{\dag}-2\bm{1}_2)
    + \textrm{sin}\beta \cdot (\textrm{cos}(2\eta/f_{\eta})-1)
    \right] \label{general_beta_potential} \,.
\end{gather}

In order to construct the ground state, we set $\pi_{1,2}=0$ without the loss of generality due to the $\SU(2)_{\rm V}$ vector symmetry.
Furthermore, we introduce dimensionless fields $\phi_{0,3}$ defined as $\phi_0\equiv \eta/f_{\eta}$ and $\phi_3\equiv \pi_3/f_{\pi}$, respectively, and the dimensionless variables,
\begin{gather}
    \zeta \equiv \frac{\sqrt{C}z}{f_{\eta}} \,, \qquad
    \epsilon \equiv 1 - \left(\frac{f_{\pi}}{f_{\eta}} \right)^2 \,, \qquad
    S \equiv \frac{\Omega \mu_{\textrm{B}}}{2\pi^2 N_{\textrm{c}}f_{\eta}\sqrt{C}} \label{dimensionless_variables} \,.
\end{gather}
\begin{figure}[tb]
    \centering
    \includegraphics[width=15.0cm]{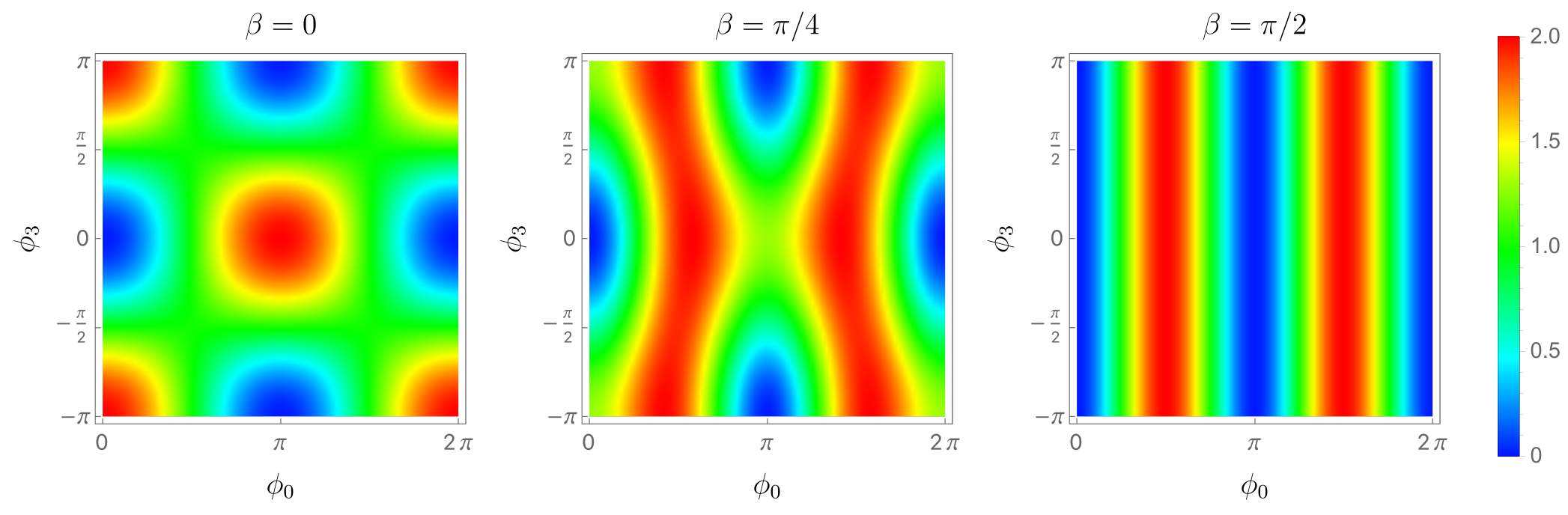}
    \caption{
    The density plot of the potential $V_{\textrm{pot}}/C$ at $\beta=0, \pi/4, \pi/2$, respectively.
    The horizontal lines of these graphs are $\phi_3$, and the vertical lines are $\phi_0$.
    As $\beta$ increases, $\phi_3$-direction of $V_{\textrm{pot}}$ flattens.
    }
    \label{fig:potential}
\end{figure}
Then, eq.~(\ref{general_beta_potential}) reduces to
\begin{gather}
    \frac{V_{\textrm{pot}}}{C} = \cos \beta (1-\cos \phi_0 \, \cos \phi_3)
    + \sin \beta (1-\cos 2\phi_0) \label{eq:V_pot} \,,
\end{gather}
which is sketched in fig.~\ref{fig:potential}.
The potential is positive semi-definite because of the condition $0\le \beta \le \pi/2$. Hence, the potential minimum is $V_{\rm pot} = 0$.
For $0\le \beta<\pi/2$
there are discrete minimum points at $(\phi_0,\phi_3)=(2n\pi,2m\pi)$ and $((2n+1)\pi ,(2m+1)\pi )$ with $n,m \in \mathbb{Z}$. 
Equivalently, these can also be represented as $(e^{i\phi_0},\Sigma=e^{i\tau_A\phi_A}) = (1,{\bf 1}_2)$ and $(-1,-{\bf 1}_2)$.
However, all these vacua are identical since they correspond to $U={\bf 1}_2$.
The equivalence between $(1,{\bf 1}_2)$ and $(-1,-{\bf 1}_2)$ is nothing but the $\mathbb{Z}_2$ quotient of ${\rm U}(2) \simeq \frac{{\rm U}(1)\times {\rm SU}(2)}{{\mathbb{Z}_2}}$.
The potential flattens out in the direction of $\phi_3$ as $\beta$ increases from $0$ to $\pi/2$, and it does not depends on $\phi_3$ (explicitly speaking, $V_{\rm pot}$ is independent of $\Sigma$) at $\beta = \pi/2$. Therefore, the vacua at $\beta = \pi/2$ are the discrete straight lines at $\phi_0 = m\pi$ on the $\phi_0\phi_3$ plane.

\begin{figure}[tb]
    \centering
    \includegraphics[width=13.0cm]{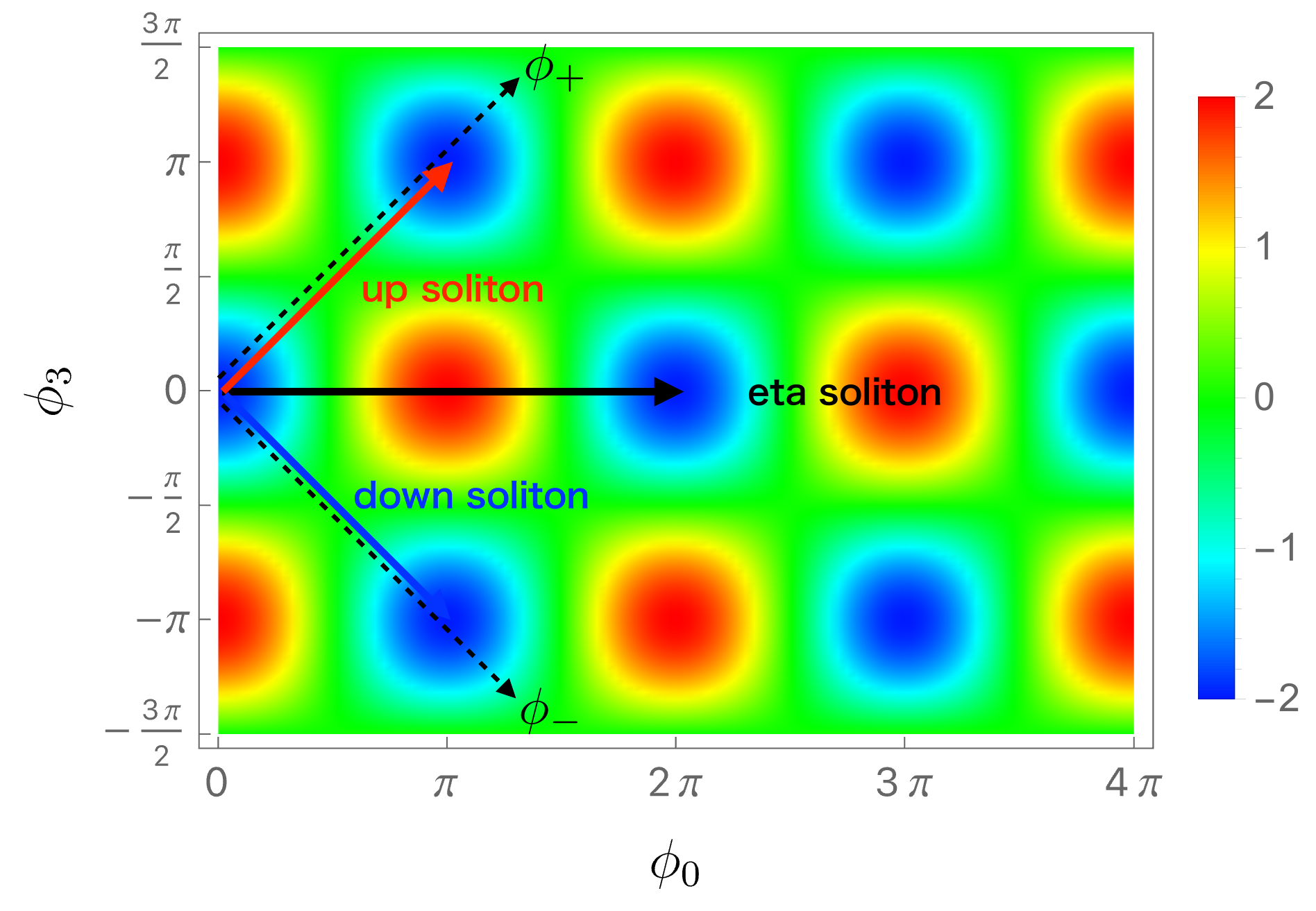}
    \caption{
    Schematic picture of an arrangement of the up (red arrow) and down (blue arrow) solitons.
    Since $\ell_+=\ell_- = \ell$, the state where the up and down solitons farthest apart satisfies $d=\ell/2$.
    This state is called the deconfined phase.
    On the other hand, the state where the up and down solitons are completely overlapped (the black arrow) is called the confined phase.
    The intermediate state satisfies $d<\ell/2$.
    }
    \label{fig:potential_and_kink}
\end{figure}

As we will shortly explain below, this model admits topologically nontrivial states because the first homotopy group of the target space is nontrivial as $\pi_1[{\rm U}(2)] \simeq \mathbb{Z}$. A minimal loop is a U$(1)$ orbit generated not by 
$\tau_0$ but by the generator $\tau_+ = (\tau_0 + \tau_3)/2$. Namely, $U = {\rm e}^{i\theta \tau_+} = {\rm diag}\left({\rm e}^{i\theta },\,1\right)$ with $\theta \in [0,2\pi]$. The loop corresponds to the red arrow connecting $(\phi_0,\phi_3) = (0,0)$ and $(\pi,\pi)$ in fig.~\ref{fig:potential_and_kink} which is not homotopic to the trivial loop $U= {\bf 1}_2$.
$U=e^{i\tau_0\theta}$ is also a nontrivial loop
corresponding to the black arrow between $(0,0)$ and $(2\pi,0)$ in fig.~\ref{fig:potential_and_kink} but this is not minimal because it winds twice about the target space $U(2)$. It is important to realize that the minimal loop is infinitely degenerate because $U' = V{\rm e}^{i\tau_+\theta}V^\dag$ with an arbitrary $V\in {\rm SU}(2)$ obviously has the same property as $U = {\rm e}^{i\tau_+\theta}$. 
For example, if we take $V = i\tau_1$, we have $U' = e^{i\tau_-\theta}$ with $\tau_- = (\tau_0-\tau_3)/2$ which corresponds to the blue arrow between $(0,0)$ and $(\pi,-\pi)$ in fig.~\ref{fig:potential_and_kink}.

Accordingly, kinks exist corresponding to the nontrivial loops. 
Let us call the kinks with $e^{i\theta \tau_+}$ and $e^{i\theta \tau_-}$ an up and down soliton, respectively.
They are continuously connected by an $\SU(2)_{\textrm{V}}$ transformation as explained above.
An important property of this soliton is that it has not only a translational moduli but also the $S^2$ moduli.
To illustrate this, let us consider an up soliton $U = e^{i\tau_+\theta}$. The vacua ($U={\bf 1}_2$) are $\theta = 0$ and $2\pi$, and the kink center has $\theta = \pi$ ($U=-\tau_3$).
Therefore, the $\SU(2)$ symmetry is spontaneously broken up to the $\U(1)$ one only around the up soliton. Therefore, this soliton has the localized moduli of $\SU(2)/\U(1) \simeq S^2$.
In this way, the kink has internal degrees of freedom and is different from the usual sine-Gordon kink.

The kink itself naturally has a spatial dependence
and usually has more energy than the vacuum state.
However, in this case,
the fifth term of eq.~(\ref{lagrangian}) reduces the energy of the modulating coordination in the $z$ direction.
In the following,
we consider only the dependence in the $z$ direction in order to investigate the possibility that a soliton appears in the ground state.
Hence, the reduced Hamiltonian density is calculated as
\begin{gather}
    \frac{\mathcal{H}}{C}
    = \frac{1-\epsilon}{2}(\phi_3^\prime)^2 + \frac{1}{2}(\phi_0^\prime)^2
    + \sin{\beta} (1-\cos{2\phi_0}) + \cos{\beta} (1-\cos{\phi_0}\cos{\phi_3}) - S \phi_0^{\prime} \label{effective_hamiltonian} \,,
\end{gather}
where the prime denotes the differentiation with respect to $\zeta$.

It is useful to change the basis by
\begin{gather}
    \phi_{\pm} \equiv \phi_0 \pm \phi_3 \label{basis_pm} \,.
\end{gather}
Using these bases, 
the Hamiltonian can be written as
\begin{eqnarray}
    \frac{\mathcal{H}}{C}
    &=& \frac{1}{2}\left[\frac{1-\epsilon/2}{2}(\phi_+^\prime)^2 + \cos{\beta} (1-\cos\phi_+) - S \phi_+^\prime \right] \notag\\
    && +\, \frac{1}{2} \left[ \frac{1-\epsilon/2}{2}(\phi_-^\prime)^2 + \cos{\beta} (1-\cos\phi_-) - S \phi_-^\prime \right] \notag\\
    && +\, \frac{\epsilon}{4}\phi_+^{\prime}\phi_-^{\prime}
    + \sin{\beta} (1-\cos(\phi_++\phi_-))\,,
    \label{effective_hamiltonian2}
\end{eqnarray}
where have used the following equation 
\begin{gather}
    -\frac{\epsilon}{2}(\phi_3^{\prime})^2
    = -\frac{\epsilon}{8}(\phi_+^{\prime})^2 - \frac{\epsilon}{8}(\phi_-^{\prime})^2
    + \frac{\epsilon}{4}\phi_+^{\prime}\phi_-^{\prime} \label{epsilon_potential} \,.
\end{gather}

\subsection{The noninteractive case of $\epsilon=0$ and $\beta=0$}

In this subsection, we consider 
the simplest case of $\epsilon=0$ and $\beta=0$.
The Hamiltonian in eq.~(\ref{effective_hamiltonian2}) is reduced 
in this case to 
\begin{gather}
    \frac{\mathcal{H}}{C}
    = \frac{1}{2}\left[
    \frac{\left(\phi_+^\prime \right)^2}{2}
    + (1-\cos{\phi_+}) - S\phi_+^\prime
    \right]
    + \frac{1}{2}\left[
    \frac{\left(\phi_-^\prime \right)^2}{2}
    + (1-\cos{\phi_-}) - S \phi_-^\prime
    \right] \,.
    \label{effective_hamiltonian3}
\end{gather}
Therefore, $\phi_{\pm}$ are completely decoupled, and each Hamiltonian is a half of that of $\phi_0$ (see eq.~(\ref{effective_hamiltonian}) at $\epsilon=0$ and $\beta=0$).
The equation of motion of $\phi_{\pm}$ is
\begin{gather}
    \phi_{\pm}^{\prime\prime} = \sin{\phi_{\pm}} \,,
\end{gather}
which can be analytically solved by using the Jacobi's elliptic function as
\begin{gather}
    \phi_{\pm} = 2 \textrm{am}\left(\frac{\zeta-\zeta_{\pm}}{k_{\pm}} \right) + \pi \label{sol_jacobi_am} \,,
\end{gather}
where $k_{\pm}\, (0<k_{\pm}<1)$ is the elliptic modulus,
and $\zeta_{\pm}$ are translational moduli of the up and down CSLs.
The period of each solution is given by
\begin{gather}
    \ell_{\pm} = 2k_{\pm}K(k_{\pm}) \,,
    \label{period}
\end{gather}
where $K(k)$ is the complete elliptic integral of the first kind.
The elliptic moduli $k_\pm$ are determined by an energy minimization condition as follows.
Substituting eq.~(\ref{sol_jacobi_am}) into the Hamiltonian in eq.~(\ref{effective_hamiltonian3}),
we can calculate the CSL energy for one period
which depends on $k_{\pm}$.
The minimization condition of this energy turns out to be
\begin{gather}
    \frac{E(k_{\pm})}{k_{\pm}} = \frac{\pi S}{4} \label{minimization_condition_of_csl} \,,
\end{gather}
where $E(k_{\pm})$ is the complete elliptic integral of the second kind.
This determines $k_\pm$ for a given angular velocity $S$. Combining this and eq.~(\ref{period}), one can get $\ell_\pm$ as a function of $S$ with $k_\pm$ being the parameters.
Since the elliptic integral satisfies the inequality $E(k_{\pm})/k_{\pm} \ge 1$, we are led to the condition $S \ge 4/\pi$. Hence the critical angular velocity $S_{\textrm{c}}$ is given by
\begin{gather}
    S_{\textrm{c}} = \frac{4}{\pi} \,.
\end{gather}
Since the right-hand side of eq.~(\ref{minimization_condition_of_csl}) of the up and down solitons are respectively equal, we show $k_+=k_-$ and $\ell_+=\ell_- \equiv \ell$.
We emphasize that the CSL state is energetically more stable than the vacuum state $\phi_{\pm}=0$ when $S \ge S_{\rm c}$ is satisfied.

The relative distance $d$ is a free parameter for $\epsilon = \beta = 0$.
On the other hand,
when we turn on $\epsilon$ and $\beta$, $d$ is no longer free to be chosen since the up and down solitons interact with each other. The distance $d$ is dynamically determined.
The states with $d=\ell/2$, $d<\ell/2$, and $d=0$ are called
a deconfined, dimer, and confined phases, respectively (see fig.~\ref{fig:phases}).

\subsection{Confined phase}
\begin{figure}[tb]
    \centering
    \includegraphics[width=15.0cm]{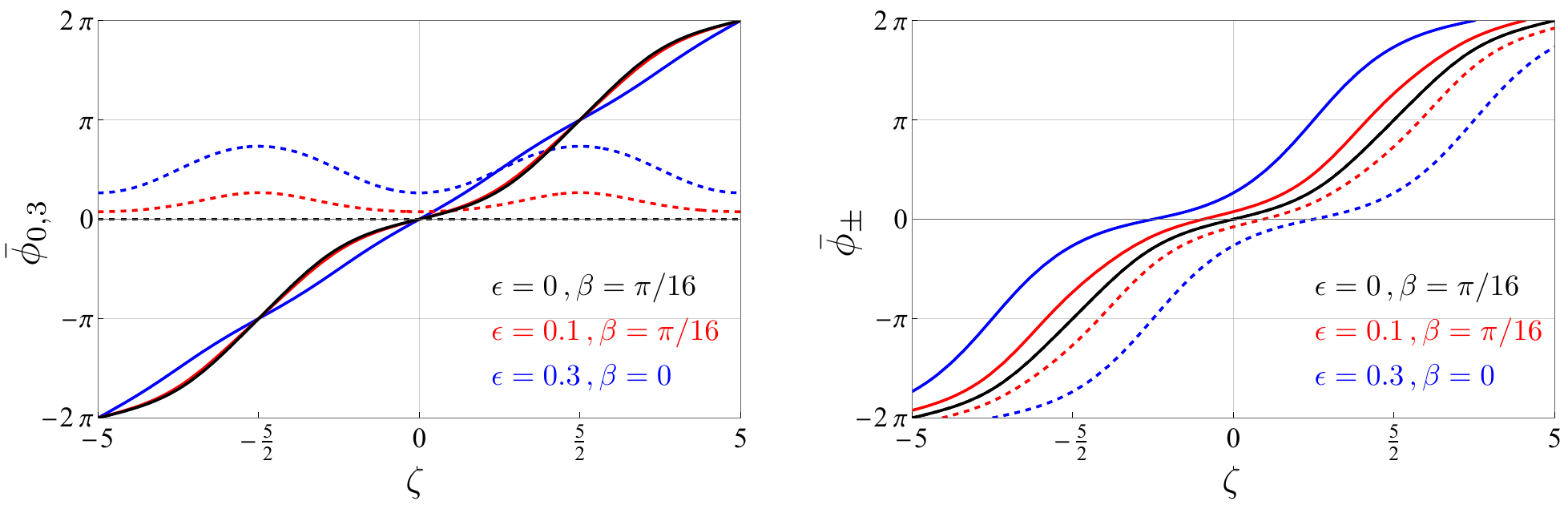}
    \caption{
    Numerical solution of $\chi_{0,3}$ (left panel) and $\chi_{\pm}$ (right panel) for $\ell=5$.
    The black, red, and blue lines correspond to the confined, dimer, and deconfined phases, respectively.
    The solid (dashed) curves show $\chi_0 (\chi_3)$ in the left panel,
    and $\chi_+ (\chi_-)$ in the right one.
    }
    \label{fig:conf}
\end{figure}
In this section, we first explain the effects of the term depending on $\epsilon>0$.
Since the kink solution of $\phi_{\pm}$ has the leak at the center of the kink,
the interaction between up and down kinks is repulsive due to the third term in eq.~(\ref{epsilon_potential}).
The $\epsilon$-dependence of $\phi_{\pm}$ is described in the left panel of fig.~\ref{fig:conf}.
As $\epsilon$ increases, the distance between $\phi_+$ and $\phi_-$ also increases.

We next explain the effects of the third and fourth terms in eq.~(\ref{effective_hamiltonian}).
This potential is sketched in fig.~\ref{fig:potential}.
As $\beta$ increases, the $\phi_3$ direction becomes flat.
Therefore, modulation in $\phi_3$ is suppressed at small $\epsilon$ (large $\beta$) (see the left panel of fig.~\ref{fig:conf}).

Therefore, a non-zero positive $\epsilon$ yields a repulsive force between the up and down solitons, whereas a non-zero positive $\beta$ leads to an attractive inter-soliton force.
When $\beta$ is sufficiently larger than $\epsilon$ such that the attraction dominates over the repulsion,
the up and down solitons overlap entirely, then $d=0$.
This phase corresponds to the confined phase (see the third line of fig.~\ref{fig:phases}).
This soliton is nothing but an $\eta$ soliton ($\phi_3$ is identically zero)
not spontaneously breaking the isospin symmetry ${\rm SU(2)_V}$ and thus is called an Abelian soliton.

\subsection{Deconfined phase}
We next consider the case where the repulsive force is sufficiently large that the up and down solitons separate as far apart as possible.
In this situation, the up and down solitons alternatively align with equal distance $d=\ell/2$ as sketched in the first line of fig.~\ref{fig:phases}.
One of the important differences between the confined and deconfined phases is that not only $\phi_0$ but also $\phi_3$ have the nontrivial kink profiles in the latter case.
Therefore, the non-Abelian part (or $\Sigma \in \SU(2)$) has a nontrivial expectation value, and then this state is called the non-Abelian CSL.
The other is that the isospin symmetry $\SU(2)_{\textrm{V}}$ is spontaneously broken down to $\U(1)$.
Hence, 
the NG modes 
associated with this spontaneous symmetry breaking  
appear in the vicinity of the soliton.
Consequently, each soliton has the moduli $S^2 \simeq \SU(2)_{\textrm{V}}/\U(1)$. 
We also note that the deconfined phase has a discrete symmetry under the $d$ shift in the $z$ direction, replacing the up (down) soliton with the down (up) one, as expressed by 
\begin{gather}
    z \to z - d \,, \qquad
    \tau_1 \textrm{diag}(\phi_+, \phi_-)\tau_1 = \textrm{diag}(\phi_-, \phi_+) \label{discrete_symm} \,.
\end{gather}

\subsection{Dimer phase}
Finally, let us consider the case where attractive and repulsive forces are balanced.
In this situation, the up and down solutions are neither completely separate nor completely overlapping.
The up and down solitons form a molecule whose size $d$ satisfies $0 < d<\ell/2$.
Then, we call this state the dimer phase.
The isospin symmetry $\SU(2)_{\textrm{V}}$ is spontaneously broken down to $\U(1)$ as in the deconfined phase.
Therefore, it is noteworthy that the isospin also appears as the NG bosons.
On the other hand, the dimer phase does not have the discrete symmetry (\ref{discrete_symm}) because of $\ell/2 \neq d$.

\section{Dispersion relations around chiral soliton lattices}
\label{sec:dispersion}

The CSL is the one-dimensional crystal of the kinks.
Similar to an ordinary crystal of atoms, quasi particles like phonons propagate through the CSL. The Abelian CSL of the $\eta$ meson is essentially identical to the well-known soliton lattice in the sine-Gordon model, and the dispersion relation of excitation modes was clarified in Ref.~\cite{Sutherland:1973zz}.
In contrast, the non-Abelian CSL includes a rich spectrum.
In this section, we study fluctuations around 
the CSL phases to numerically  obtain 
dispersion relations 
for NG modes and gapful modes.
After deriving equations for fluctuations in the background of the CSLs, 
we study 
the noninteractive case of 
$\epsilon=0$ and $\beta=0$, 
confined phase ($\eta$-CSL), 
deconfined phase 
and dimer phase 
in each subsection.

\subsection{Equations of motion for fluctuations}
In order to calculate the dispersion relations of the NG modes in the non-Abelian CSLs,
we approximate eq.~(\ref{lagrangian}) up to the quadratic of fluctuations from the background configurations $\{\bar\phi_0,\bar\phi_3\}$ with the following boundary condition:
\begin{gather}
{\bar \phi}_0(0)=0 \,, \qquad {\bar \phi}_0(\ell)=2\pi \,, \label{bc_0} \\
{\bar \phi}^{\prime}_3(0) = {\bar \phi}^{\prime}_3(\ell)=0 \,, \label{bc_3}
\end{gather}
where $\ell$ is a period of the kinks, see fig.~\ref{fig:conf} (a).
We introduce small fluctuations as follows:
\begin{gather}
    \Sigma = e^{\rmi \tau_3 \bar{\phi}_3} \delta \Sigma \,, \qquad
    \eta/f_\eta = \phi_0 = {\bar \phi}_0 + \delta  \phi_0 \,
\end{gather}
where the fluctuation of the $\SU(2)$ part satisfies $\det \delta \Sigma=1$ due to the condition $\det \Sigma=1$.
Then, we can represent $\delta \Sigma$ as
\begin{gather}
    \delta \Sigma = \sqrt{1-\delta \phi_A^2} + \rmi \tau_A\delta \phi_A \,.
\end{gather}
$\delta\Sigma \delta\Sigma^\dag = {\bf 1}_2$ is also satisfied.
Substituting the above equations into eq.~(\ref{lagrangian}), the Lagrangian of the quadratic order in the fluctuations is given by
\begin{eqnarray}
    {\cal L}^{(2)} &=& \frac{f_\eta^2}{2}(\del_\mu\delta\phi_0)^2 + \frac{f_\pi^2}{2}(\del_\mu\delta\phi_A)^2 - f_\pi^2 \epsilon_{3AB}\del_\mu\bar\phi_3 \delta\phi_A\del^\mu\delta\phi_B \notag\\
    &&+\, \frac{C \cos\beta}{2}\left( - \delta\phi_a^2 \cos\bar\phi_0 \cos\bar\phi_3  + 2 \delta\phi_0\delta\phi_3 \sin\bar\phi_0 \sin\bar\phi_3 \right) \notag\\
    &&-\, 2 C \sin\beta \cos2\bar\phi_0\,\delta\phi_0^2\,,
\end{eqnarray}
This can be separated into two decoupled portions as
\begin{eqnarray}
    {\cal L}^{(2)}_{\rm neutral} &=& \frac{f_\eta^2}{2}(\del_\mu\delta\phi_0)^2 + \frac{f_\pi^2}{2}(\del_\mu\delta\phi_3)^2 -\, 2 C \sin\beta \cos2\bar\phi_0\,\delta\phi_0^2 \notag\\
    &&+\, \frac{C \cos\beta}{2}\left[ - \left(\delta\phi_0^2+\delta\phi_3^2\right) \cos\bar\phi_0 \cos\bar\phi_3  + 2 \delta\phi_0\delta\phi_3 \sin\bar\phi_0 \sin\bar\phi_3 \right]\,,\\
    {\cal L}^{(2)}_{\rm charged} &=& \frac{f_\pi^2}{2}\left[(\del_\mu\delta\phi_1)^2+(\del_\mu\delta\phi_2)^2\right] - f_\pi^2 \epsilon_{3AB}\del_\mu\bar\phi_3 \delta\phi_A\del^\mu\delta\phi_B \notag\\
    &&-\, \frac{C \cos\beta}{2}   \cos\bar\phi_0 \cos\bar\phi_3 \left(\delta\phi_1^2 + \delta\phi_2^2\right)\,,
\end{eqnarray}

We refer the neutral components $\delta\phi_{0,3}$ to phonons because they are related to the NG and quasi-NG bosons associated with the spontaneously broken translational symmetry by the CSL. On the other hand, we call the charged fluctuations $\delta\phi_{1,2}$ as the isospinons \cite{Eto:2021gyy} are the NG bosons because they are associated with the spontaneously broken isospin symmetry $SU(2)_{\rm V}$ by the background $\bar\phi_3 \neq 0$.

\subsubsection{The neutral components: phonons}
The linearized EOM of $\delta \phi_0$ is given by
\begin{gather}
    \left(\del_{\bar t}^2-\del^2_{{\zeta}}\right)\delta \phi_0
    + \left(
    4 \sin \beta \cos 2\bar \phi_0 + \cos \beta \cos\bar \phi_0 \cos\bar \phi_3
    \right)\delta \phi_0
    - \left(\cos \beta  \sin\bar \phi_0 \sin\bar \phi_3 \right) \delta\phi_3 = 0 \label{fluc_delta_phi} \,, 
\end{gather}
where we have used dimensionless variables in eq.~(\ref{dimensionless_variables}), the dimensionless coordinate defined as ${\bar t} \equiv \sqrt{C}t/f_{\eta}$.
Similarly, the linearized EOM for $\delta\phi_3$ reads
\begin{gather}
        (1-\epsilon)\left(\del_{\bar t}^2-\del^2_{{\zeta}}\right)\delta \phi_3 - \left(\cos\beta \sin\bar \phi_0 \sin\bar\phi_3\right) \delta \phi_0 
    + \left(\cos\beta \cos \bar \phi_0 \cos \bar\phi_3\right) \delta \phi_3 = 0 \label{fluc_delta_phi3} \,.
\end{gather}
Note that we have assumed the fluctuations depend only on $t$ and $z$ for simplicity.\footnote{Including $x^1$ and $x^2$ makes the fluctuation analysis much more complicated due to the rotation effect in the metric.}

We first focus on the neutral components $\phi_0$ and $\phi_3$.
It turns out that the following redefinition of the fields is useful
\begin{gather}
    \delta \phi_+ \equiv \delta \phi_0 + \delta \phi_3 \,, \qquad
    \delta \phi_- \equiv \delta \phi_0 - \delta \phi_3 \,.
\end{gather}
In terms of these new fields, together with the decomposition
\begin{eqnarray}
\delta \phi_{\pm}=\rme^{-\rmi \bar \omega \bar t}\varphi_{\pm}(\zeta)\,,    
\end{eqnarray} 
the linearized EOMs reduce 
\begin{eqnarray}
\left[
- \del_\zeta^2 + \left(
\begin{array}{cc}
P_+ & Q_-\\
Q_+ & P_-
\end{array}
\right)
\right]
\left(
\begin{array}{c}
\varphi_+\\
\varphi_-
\end{array}
\right)
= \bar \omega^2 
\left(
\begin{array}{c}
\varphi_+\\
\varphi_-
\end{array}
\right)\,,
\label{EOM of varphi+-}
\end{eqnarray}
with
\begin{eqnarray}
    P_\pm &=& 2\sin\beta \cos 2\bar\phi_0
    + \frac{2-\epsilon}{2(1-\epsilon)} \cos\beta \cos\bar \phi_\pm\,,\\
    Q_\pm &=& 2\sin\beta \cos2{\bar \phi}_0
    - \frac{\epsilon}{2(1-\epsilon)} \cos\beta \cos\bar \phi_\pm\,.
\end{eqnarray}

In order to find a collect dispersion relation, we should take the periodicity of the potential into account. The common periodicity in $\zeta$ among $\cos2\bar\phi_0$ and $\cos\bar\phi_3$ is the lattice constant $\ell$, so that $P_\pm$ and $Q_\pm$ have the same periodicity.
Hence, this equation has a form similar to the Schr$\ddot{\textrm{o}}$dinger equation for electrons in a crystal.
Therefore, we can apply the Bloch theorem to our system, and $\varphi_{\pm}(\zeta)$ satisfies the following equation:
\begin{gather}
    \varphi_{\pm}(\zeta+\ell) = \rme^{\rmi \bar p \ell}\varphi_{\pm}(\zeta) \label{eq:Bloch_theorem} \,,
\end{gather}
where $\bar p$ is a dimensionless crystal momentum vector.
We consider the periodic boundary condition, which is the so-called Born–von Karman boundary condition:
\begin{gather}
    \varphi_{\pm}(\zeta+N\ell) = \varphi_{\pm}(\zeta) \label{eq:BvK_condition}
\end{gather}
where $N$ is a positive integer.
Substituting eq.~(\ref{eq:BvK_condition}) into eq.~(\ref{eq:Bloch_theorem}),
we get $\rme^{\rmi \bar pN\ell}=1$.
As a result, the crystal momentum $\bar p$ takes discrete values as
\begin{gather}
    \bar p_n = \frac{2\pi n}{N\ell} \quad (n \in \mathbb{Z}) \label{eq:discretized_momentum} \,.
\end{gather}
In fact, it is sufficient to consider $\bar p_n$ in the following range
\begin{gather}
    -\frac{\pi}{\ell} < \bar p_n \le \frac{\pi}{\ell} \label{first_BZ} \,.
\end{gather}
This is called the first Brillouin zone which is a unit cell of the reciprocal lattice.
Note that, when we replace $p_n$ with $\bar p_n+2\pi j/\ell$ $(j \in \mathbb{Z})$,
$\varphi_{\pm}(\zeta)$ satisfies the following equation due to the Bloch theorem:
\begin{gather}
    \varphi_{\pm}(\zeta+\ell)
    = \rme^{\rmi\left(\bar p_n + g_j \right)\ell} \varphi_{\pm}(\zeta)
    = \rme^{\rmi \bar p_n \ell}\varphi_{\pm}(\zeta) \,,
    \label{eq:reciprocal_lattice_shift}
\end{gather}
where we define $g_j\equiv 2\pi j/\ell$ corresponding to the reciprocal lattice vector.
Hence, this momentum shift simply leads back to the same wave function.
This implies that there are multiple $\varphi_{\pm}(\zeta)$ for the same $\bar p_n$ in the first Brillouin zone.
We will use the label $j$ to distinguish them
(\( j \) is a label specifying the band structure).
For later convenience,
we define $u^{(n)}_{\pm}(\zeta)$ as follow
\begin{gather}
    \varphi^{(n)}_{\pm}(\zeta) = \rme^{\rmi \bar p_n \zeta} u^{(n)}_{\pm}(\zeta) \label{eq:bloch_wavefunction} \,,
\end{gather}
where we impose 
$u^{(n)}_{\pm}(\zeta+\ell) = u^{(n)}_{\pm}(\zeta)$.
Inserting eq.~(\ref{eq:bloch_wavefunction}) into eq.~(\ref{EOM of varphi+-}),
we get the Sch\"odineger equation for $u_\pm^{(n)}$
\begin{gather}
    \left[
    -(\del_{\zeta}+\rmi \bar p_n)^2
    + \left(
    \begin{array}{cc}
    P_+ & Q_-\\
    Q_+ & P_-
    \end{array}
    \right)
    \right] 
    \left(
    \begin{array}{c}
    u^{(n)}_{+} \\
    u^{(n)}_{-}
    \end{array}
    \right)
    = \bar{\omega}_{(n,j)}^2
    \left(
    \begin{array}{c}
    u^{(n)}_{+} \\
    u^{(n)}_{-}
    \end{array}
    \right) \label{eq:EOM_of_u+-} \,.
\end{gather}

Below, we will attempt to numerically solve eq.~(\ref{eq:EOM_of_u+-}).
To this end, it is convenient to perform the Fourier expansion of eq.~(\ref{eq:EOM_of_u+-}).
Since $u^{(n)}_{\pm}$ has a period $\ell$,
it can be expanded by the Fourier series as
\begin{gather}
    u^{(n)}_{\pm}(\zeta)
    = \sum_m f^{(n;m)}_{\pm}\rme^{\rmi \frac{2\pi m}{\ell}\zeta}
    = \sum_m f^{(n;m)}_{\pm}\rme^{\rmi g_m\zeta} \,.
\end{gather}
Then, the derivative term is expanded as
\begin{gather}
    -(\del_{\zeta}+\rmi \bar p_n)^2u^{(n)}_{\pm}(\zeta)
    = \sum_m (\bar p_n+g_m)^2f^{(n;m)}_{\pm}\rme^{\rmi g_m\zeta} \,.
\end{gather}
Similarly, we expand the background functions $\bar\phi_0$ and $\bar \phi_\pm$ and express them as
\begin{eqnarray}
    \cos \bar{\phi}_{\pm} &=& \sum_{n}c_\pm^{(n)} \rme^{\rmi \frac{2\pi n}{\ell}\zeta} = \sum_{n}c_\pm^{(n)} \rme^{\rmi g_n\zeta} \label{eq:Fourier_cos_phi_pm} \,,  \\
    \cos 2\bar\phi_0 &=& \sum_{n}c_0^{(n)}\rme^{\rmi \frac{2\pi n}{\ell} \zeta} = \sum_{n}c_0^{(n)}\rme^{\rmi g_n\zeta} \,.
\end{eqnarray}
So, for example, the Fourier expansion of  
$u^{(n)}_{\pm} \cos \bar{\phi}_+$ is given by
\begin{align}
    u^{(n)}_{\pm}\cos \bar{\phi}_{\pm}
    = \sum_{m,m'} f_{\pm}^{(n;m)} c_\pm^{(m')}\rme^{\rmi (g_m+g_{m'})\zeta}
    = \sum_m\left(
    \sum_{m^{\prime}} c^{(m^{\prime})}_{\pm} f^{(n;m-m^{\prime})}_{\pm}
    \right)\rme^{\rmi g_{m}\zeta}\,.
\end{align}
Similar expressions can be easily obtained for the other terms
$u^{(n)}_{\pm} \cos2\bar\phi_0$.
Then, we get the Fourier series of the left-hand side of eq.~(\ref{EOM of varphi+-})
\begin{eqnarray}
    &-& \frac{d^2}{d\zeta^2}u^{(n)}_{+} + P_+ u^{(n)}_{+} + Q_- u^{(n)}_{-} \notag \\
    &&= \sum_m \rme^{\rmi  g_n\zeta} \left\{ 
    (\bar{p}_n+g_m)^2 f_+^{(n;m)}
    + \sum_{m^{\prime}} \left( A_+^{(m^{\prime})} f_+^{(n;m-m^{\prime})}
    + B_-^{(m^{\prime})} f_-^{(n;m-m^{\prime})}\right) \right\}\,,\\
    &-& \frac{d^2}{d\zeta^2}u^{(n)}_{-} + Q_+ u^{(n)}_{+}  + P_- u^{(n)}_{-}  \notag \\
    &&= \sum_m \rme^{\rmi g_n\zeta} \left\{ 
    (\bar{p}_n+g_m)^2 f_-^{(n;m)}
    + \sum_{m^{\prime}}  \left( B_+^{(m^{\prime})} f_+^{(n;m-m^{\prime})} 
    + A_-^{(m^{\prime})} f_-^{(n;m-m^{\prime})}\right) \right\}\,.
\end{eqnarray}
with
\begin{eqnarray}
    A_\pm^{(m)} &\equiv& 2 c_0^{(m)} \sin\beta + \frac{2-\epsilon}{2(1-\epsilon)}c_\pm^{(m)} \cos\beta\,,\\
    B_\pm^{(m)} &\equiv& 2 c_0^{(m)} \sin\beta - \frac{\epsilon}{2(1-\epsilon)} c_\pm^{(m)} \cos\beta\,.
\end{eqnarray}
Plugging all these into eq.~(\ref{EOM of varphi+-}), we eventually find 
\begin{eqnarray}
    \bar\omega_{(n)}^2 f_+^{(n;m)} &=& (\bar p_n + g_m)^2 f_+^{(n;m)}
    + \sum_{m^{\prime}} \left( A_+^{(m^{\prime})} f_+^{(n;m-m^{\prime})}
    + B_-^{(m)} f_-^{(n;m-m^{\prime})}\right)\,,\label{eq:f+}\\
    \bar \omega_{(n)}^2 f_-^{(n;m)} &=& (\bar p_n + g_m)^2 f_-^{(n;m)}
    + \sum_{m^{\prime}}  \left( B_+^{(m^{\prime})} f_+^{(n;m-m^{\prime})} 
    + A_-^{(m^{\prime})} f_-^{(n;m-m^{\prime})}\right)\,.\label{eq:f-}
\end{eqnarray}
One can see that $f_\pm^{(n;m)}$ couples only to $f_\pm^{(n';m^{\prime})}$ with $n' = n$.
Composing $f_\pm^{(n;m)}$ into the following vectors
\begin{eqnarray}
    \bm{f}_{\pm}^{(n)}
    \equiv \left(\cdots,f_{\pm}^{(n;2)},f_{\pm}^{(n;1)},f_{\pm}^{(n;0)},f_{\pm}^{(n;-1)},f_{\pm}^{(n;-2)},\cdots \right)^{\rm T} \,,
\end{eqnarray}
eqs.~(\ref{eq:f+}) and (\ref{eq:f-}) are cast into the matrix equation
\begin{eqnarray}
    \bar\omega_{(n)}^2
    \left(
    \begin{array}{c}
    \bm{f}^{(n)}_+\\
    \bm{f}^{(n)}_-
    \end{array}
    \right)
    =
    \left(
    \begin{array}{cc}
    C_+^{(n)} & D_-\\
    D_+ & C_-^{(n)}
    \end{array}
    \right)
    \left(
    \begin{array}{c}
    \bm{f}^{(n)}_+\\
    \bm{f}^{(n)}_-
    \end{array}
    \right)\,, \label{eq:eigen_eq}
\end{eqnarray}
with
\begin{eqnarray}
    C_\pm^{(n)} &\equiv& 
    \left(
    \begin{array}{ccccc}
    \ddots & & & & \\
    & (\bar p_n + g_{-1})^2 & & & \\
    & & \bar p_n^2 & & \\
    & & & (\bar p_n + g_{1})^2 & \\
    & & & & \ddots
    \end{array}
    \right)
    +
    \left(
    \begin{array}{ccccc}
    & \vdots & \vdots & \vdots & \\
    \cdots &  A_\pm^{(0)} & A_\pm^{(-1)} & A_\pm^{(-2)}&  \cdots\\
    \cdots &  A_\pm^{(1)} & A_\pm^{(0)} & A_\pm^{(-1)} &  \cdots \\
    \cdots &  A_\pm^{(2)} & A_\pm^{(1)} & A_\pm^{(0)} &  \cdots \\
    & \vdots & \vdots & \vdots & 
    \end{array}
    \right)\,,\\
    D_\pm &\equiv&
\left(
    \begin{array}{ccccc}
    & \vdots & \vdots & \vdots & \\
    \cdots &  B_\pm^{(0)} & B_\pm^{(-1)} & B_\pm^{(-2)}&  \cdots\\
    \cdots &  B_\pm^{(1)} & B_\pm^{(0)} & B_\pm^{(-1)} &  \cdots \\
    \cdots &  B_\pm^{(2)} & B_\pm^{(1)} & B_\pm^{(0)} &  \cdots \\
    & \vdots & \vdots & \vdots & 
    \end{array}
    \right)\,.
\end{eqnarray}

The prescription for finding the dispersion relation is the following.
We first choose $n$ in $-N/2 < n \le N/2$ or the corresponding momentum $\bar p_n$ in the Brillouin zone. Then we solve the eigen-equation (\ref{eq:eigen_eq}), and determine the eigenvalues $\bar\omega_{(n,j)}$ and eigenvector ${\bm f}_{\pm}^{(n,j)}$. We repeat this for each $n$ in $-N/2 < n \le N/2$.
Note that for the chosen momentum $\bar p_n$, there are multiple states labeled by the new index $j$.
The label $j$ is nothing but the index associated with the band structure.
Though the sizes of matrices $C_{\pm}^{(n)}$ and $D_\pm$ are infinite,
we have to truncate them to be finite-size matrices for numerical computations.
We denote the finite-size matrices as
\begin{align}
    \left(\tilde C^{(n)}_{\pm}\right)_{ij}
    &= (\bar p_n + g_{i-d-1})\delta_{ij} + A^{(i-j)}_{\pm} \,, \\
    \left(\tilde D^{(n)}_{\pm}\right)_{ij}
    &= B^{(i-j)}_{\pm} \,,
\end{align}
where $i,j$ are indices of the truncated matrices running over $1, 2, \cdots, 2\Lambda+1$.
Upon diagonalizing the matrix,
we derive \(2(2\Lambda+1)\) eigenvalues.
These are distinguished by the label \(j\),
which denotes the band index.
For convenience in subsequent discussions,
we arrange \(j\) in the order \(j=1,2,\cdots,2(2\Lambda+1)\) such that
\begin{equation}
    |\bar \omega_{(n,j=1)}|
    \le |\bar \omega_{(n,j=2)}|
    \le |\bar \omega_{(n,j=3)}|
    \cdots
    \le |\bar \omega_{(n,j=2(2\Lambda+1))}| \,.
\end{equation}

\subsubsection{The charged components: isospinons}
The linearized EOM of $\delta \phi_1$ and $\delta \phi_2$ are given by
\begin{gather}
    f^2_{\pi} \left(\del_\mu^2\delta \phi_1 + \del_\mu^2{\bar \phi}_3 \delta \phi_2 + 2\del_{\mu}{\bar \phi}_3 \del^{\mu}\delta \phi_2\right)
    + C \cos\beta \cos\bar \phi \cos\bar \phi_3 \delta \phi_1 = 0 \,, \\
    f^2_{\pi}\left(\del_\mu^2\delta \phi_2 - \del_\mu^2{\bar \phi}_3 \delta \phi_1 - 2\del_{\mu}{\bar \phi}_3 \del^{\mu}\delta \phi_1\right)
    + C\cos\beta \cos{\bar \phi} \cos{\bar \phi}_3\, \delta \phi_2 = 0 \,.
\end{gather}
These can be combined into a complex equation
for $\delta \pi_{\pm}\equiv \delta \phi_1 \pm \rmi \delta \phi_2$ as
\begin{gather}
    f^2_{\pi}\left(\del_\mu^2\delta \pi_+ -\rmi  (\del_\mu^2{\bar \phi}_3) \delta \pi_+ -2\rmi (\del_{\mu}{\bar \phi}_3) \del^{\mu}\delta \pi_+\right)
    + C \cos\beta \cos{\bar \phi} \cos{\bar \phi}_3 \delta \pi_+ = 0 \label{EOM of pi+} \,.
\end{gather}
This can be further simplified by introducing the following field:
\begin{gather}
    \delta \tilde{\pi}_+ \equiv \rme^{-\rmi {\bar \phi}_3}\delta \pi_+ \,.
\end{gather}
Eq.~(\ref{EOM of pi+}) is rewritten as
\begin{gather}
    \del_\mu^2 \delta \tilde{\pi}_+
    - (\del_z{\bar \phi}_3)^2\delta \tilde{\pi}_+
    + \frac{C}{f^2_{\pi}} \cos\beta \cos{\bar \phi}_0 \cos{\bar \phi}_3 \delta \tilde{\pi}_+ = 0 \,.
\end{gather}
Using the dimensionless variables and decomposing the fluctuation as
\begin{eqnarray}
\delta \tilde{\pi}_+ = \rme^{-\rmi \bar t \bar \omega}\psi_+(\zeta)\,,    
\end{eqnarray}
we finally obtain
\begin{gather}
    \left\{ -\del^2_{{\zeta}} 
    - (\del_{{\zeta}}{\bar \phi}_3)^2
    + \frac{\cos\beta}{1-\epsilon} \cos{\bar \phi}_0 \cos{\bar \phi}_3 \right\} \psi_+ = \bar\omega^2 \psi_+ \label{EOM_of_tilde_pi+} \,.
\end{gather}

The potential is periodic with the periodicity $\ell$ as before. 
So $\psi_+$ satisfies the Bloch theorem
\begin{gather}
    \psi_+(\zeta+\ell) = \rme^{\rmi \bar p \zeta}\psi_+(\zeta) \,,
\end{gather}
where $\bar p$ is a real number.
We impose the Born-Von Karman boundary condition
\begin{eqnarray}
    \psi_+(\zeta + N\ell) = \psi_+(\zeta)\,,
\end{eqnarray}
with an integer $N$.
Then, \( \bar p \) is quantized as eq.~(\ref{eq:discretized_momentum}).
Just as in the case of \( \varphi_{\pm} \),
it is enough for us to take the momentum \( \bar p_n \) in the first Brillouin zone (\ref{first_BZ}).
Hereafter, we will denote the wave function $\psi_+^{(n)}$ with the index $n$ to specify the crystal momentum $\bar p_n$.
But, as mentioned above, there are multiple $\psi_+^{(n)}$ for the same $\bar p_n$ due to the symmetry (\ref{eq:reciprocal_lattice_shift}), which will be distinguished by the band index $j$.
In order to obtain the dispersion relation numerically, let us rewrite
$\psi_+^{(n)}$ as
\begin{gather}
    \psi_+^{(n)}(\zeta) = \rme^{\rmi \bar p_n\zeta} v_+^{(n)}(\zeta) \,,
\end{gather}
where $v_+^{(n)}(\zeta+\ell) = v_+^{(n)}(\zeta)$ is imposed.
Substituting this equation in eq.~(\ref{EOM_of_tilde_pi+}),
we get the EOM for $v_+^{(n)}$
given by
\begin{gather}
    \left[
    -(\del_{\zeta}+\rmi p_n)^2 + R(\zeta)
    \right] v_+^{(n)}
    = \bar \omega_{(n)}^2 v_+^{(n)} \,,
    \label{EOM_of_tilde_psi+_2}
\end{gather}
with
\begin{eqnarray}
    R(\zeta) \equiv - (\del_{{\zeta}}{\bar \phi}_3)^2 +
    \frac{\cos\beta}{1-\epsilon} \cos{\bar \phi}_0 \cos{\bar \phi}_3\,. 
\end{eqnarray}
Since $R(\zeta)$ 
has a period $\ell$, it is expanded as
\begin{eqnarray}
    R(\zeta) = \sum_m r_m \rme^{\rmi g_m \zeta} \,.
\end{eqnarray}
Denoting the Fourier expansion of $v_+^{(n)}$ as
\begin{gather}
    v^{(n)}_{\pm}(\zeta)
    = \sum_m g^{(n;m)}_{\pm}\rme^{\rmi 2\pi m\zeta/\ell}
    = \sum_m g^{(n;m)}_{\pm}\rme^{\rmi g_m\zeta} \,,
\end{gather}
$v_+^{(n)}R$ can be calculated as
\begin{eqnarray}
    v^{(n)}_{\pm}R
    = \sum_m\left(
    \sum_{m^{\prime}} r^{(m^{\prime})}_{\pm}
    g^{(n;m-m^{\prime})}_{\pm}
    \right)\rme^{\rmi g_{m}\zeta}
\end{eqnarray}
Then, we get the Fourier series of eq.~(\ref{EOM_of_tilde_psi+_2})
\begin{gather}
    \bar \omega_{(n)}^2g_+^{(n;m)}
    =  (\bar{p}_n+g_m)^2 g_+^{(n;m)}
    + \sum_{m^{\prime}}
    r^{(m^{\prime})} g_+^{(n;m-m^{\prime})} \,.
    \label{eq:g+}
\end{gather}
It is evident that \(g_+^{(n;m)}\) couples only with \(g_+^{(n';m')}\) with $n = n'$.
By expressing \(g_+^{(n;m)}\) as the vector
\begin{equation}
    \bm{g}_+^{(n)}
    \equiv \left(\cdots,g_{+}^{(n;2)},g_{+}^{(n;1)},g_{+}^{(n;0)},g_{+}^{(n;-1)},g_{+}^{(n;-2)},\cdots \right)^{\rm T} \,,
\end{equation}
eq.~\eqref{eq:g+} can be reformulated as a matrix equation as follows:
\begin{gather}
    \bar \omega_{(n)}^2\bm{g}_+^{(n)}
    = E^{(n)}\bm{g}_+^{(n)} \,,
\end{gather}
where the matrix $E^{(n)}$ is defined as
\begin{eqnarray}
    E^{(n)} &\equiv& 
    \left(
    \begin{array}{ccccc}
    \ddots & & & & \\
    & (\bar p_n + g_{-1})^2 & & & \\
    & & \bar p_n^2 & & \\
    & & & (\bar p_n + g_{1})^2 & \\
    & & & & \ddots
    \end{array}
    \right)
    +
    \left(
    \begin{array}{ccccc}
    & \vdots & \vdots & \vdots & \\
    \cdots &  r^{(0)} & r^{(-1)} & r^{(-2)}&  \cdots\\
    \cdots &  r^{(1)} & r^{(0)} & r^{(-1)} &  \cdots \\
    \cdots &  r^{(2)} & r^{(1)} & r^{(0)} &  \cdots \\
    & \vdots & \vdots & \vdots & 
    \end{array}
    \right)\,.
\end{eqnarray}

The prescription for the dispersion relation is the same as that for the case of the neutral components.
Firstly, we choose an integer $n$ from the Brillouin zone $-N/2 < n \le N/2$.
Then we diagonalize the matrix $E^{(n)}$ and find the eigenvalus $\bar\omega_{(n,j)}$ and the eigenvector ${\bm g}_+^{(n,j)}$.
In order to do this by numerical computations,
we have to truncate the infinitely large matrix $E^{(n)}$ to be a finite one $\tilde E^{(n)}$ whose size is $(2\Lambda+1)^2$  as before.

Now, we have completed all necessary preparations for numerically obtaining the dispersion relations for any CSL backgrounds.

\subsection{The noninteractive case of $\epsilon=0$ and $\beta=0$}
\begin{figure}[tp]
  \begin{center}
  \includegraphics[width=15.0cm]{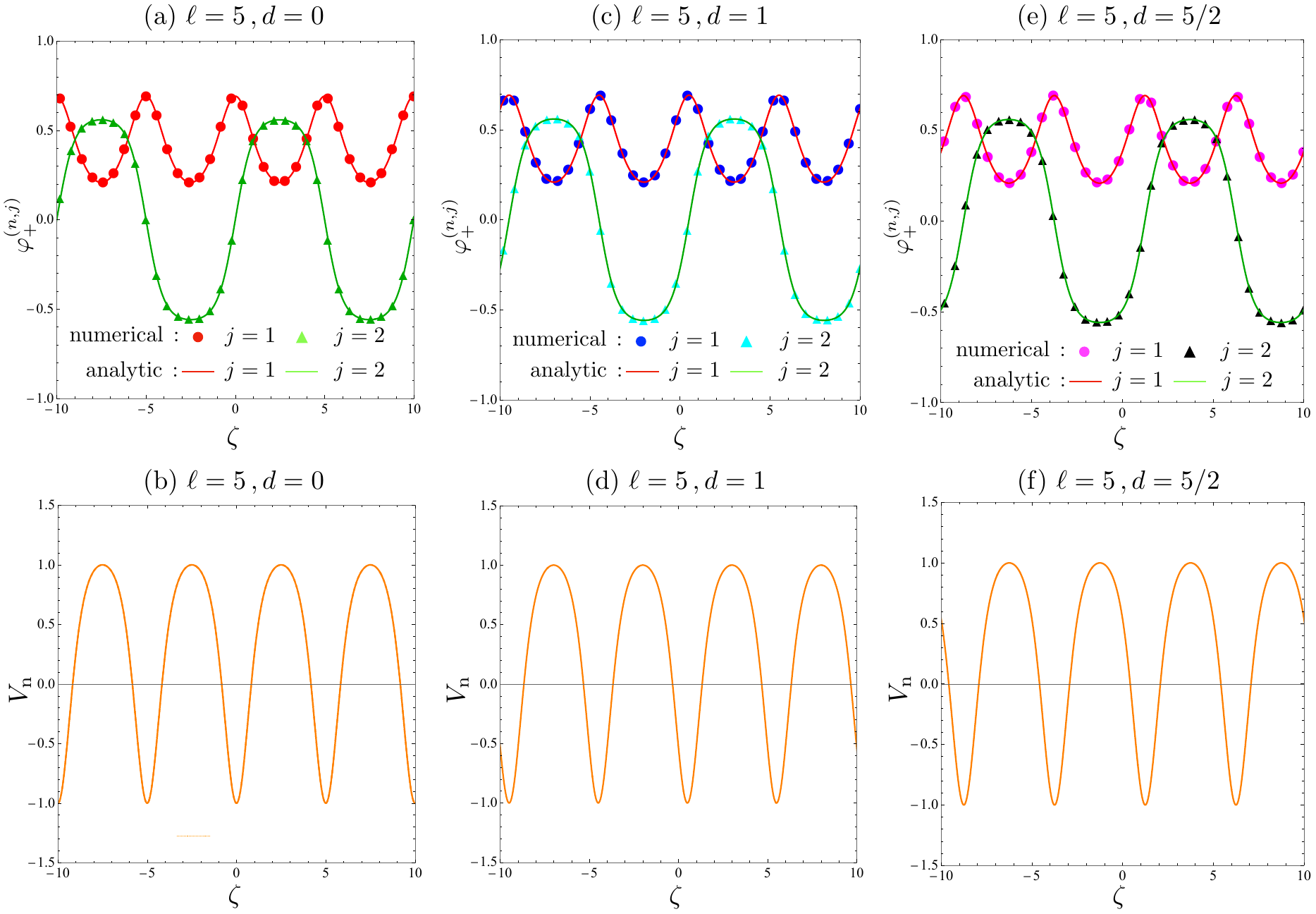}
  \end{center}
  \caption{
  [The noninteractive case: the neutral components $\varphi_+^{(n,j)}$]
  In the top row,
  $\varphi_+^{(0,1)}$ for $d=0,1,5/2$ are shown by the red, blue, and magenta-filled circles, respectively.
  $\varphi_+^{(\pi/\ell,2)}$ for $d=0,1,5/2$ are also shown by the green, cyan and black-filled triangles, respectively.
  The solid curves are the analytic results given in  eqs.~\eqref{eq:lame_massless_varphi} and \eqref{eq:lame_massless_varphi_edge}.
  In the bottom row,
  the potentials $V_{\mathrm{n}}$ for $d=0,1,5/2$ are shown.
  }
  \label{fig:wavefunction_lame_neutral}
\end{figure}

First, let us consider 
the noninteractive case of $\epsilon=0$ and $\beta=0$.
Eqs.~(\ref{EOM of varphi+-}) and (\ref{EOM_of_tilde_pi+}) become
\begin{gather}
    \left(-\del^2_{{\zeta}}
    +  \cos{\bar \phi}_{\pm}\right) \varphi_{\pm}^{(n)} = m_{(n)}^2 \varphi_{\pm}^{(n)} \label{phi_pm_at_ideal} \,, \\
    \left(-\del^2_{{\zeta}}
    - ({\bar \phi}_3^\prime)^2
    + \cos{\bar \phi}_0 \cos{\bar \phi}_3\right)\psi_+^{(n)} = \mu_{(n)}^2 \psi_+^{(n)} \label{charged_at_ideal} \,.
\end{gather}
In order to avoid a confusion in notations, we have denoted $\bar \omega_{(n)} \to m_{(n)}$ for $\varphi_{\pm}^{(n)}$, and $\bar \omega_{(n)} \to \mu_{(n)}$ for $\psi_{\pm}^{(n)}$.
The background configuration is given by
\begin{gather}
    \bar{\phi}_{\pm}
    = 2\,\textrm{am}\left(\frac{\zeta \mp d/2}{k},k \right) + \pi \,.
\end{gather}
$d$ is the relative distance between the up and down solitons which is a free parameter since there is no interaction between the up and down CSLs at $\epsilon=0$ and $\beta=0$. This case is particularly simple since $\varphi_+$ and $\varphi_-$ are decoupled.
Inserting these solutions into the potential parts of eqs.~(\ref{phi_pm_at_ideal}) and (\ref{charged_at_ideal}), we get
\begin{gather}
    \cos \bar{\phi}_{\pm} = 2\,\textrm{sn}^2_{\pm}-1 \equiv V_{\mathrm{n}} \,, \\
    - (\bar\phi_3')^2 + \cos\bar\phi_0\cos\bar\phi_3
    =  - \frac{1}{k^2}\left(
    {\rm dn}_+ - {\rm dn}_-\right)^2
    - {\rm cn}_+^2 {\rm cn}_-^2
    + {\rm sn}_+^2 {\rm sn}_-^2 \equiv V_{\mathrm{c}} \,,
\end{gather}
where we have defined $\textrm{sn}_{\pm}=\textrm{sn}((\zeta \mp d/2)/k,k)$,
$\textrm{cn}_{\pm}=\textrm{cn}((\zeta \mp d/2)/k,k)$
and $\textrm{dn}_{\pm}=\textrm{dn}((\zeta \mp d/2)/k,k)$ for short.
Thus, eqs.~(\ref{phi_pm_at_ideal}) and (\ref{charged_at_ideal}) read
\begin{gather}
    \left(
    -\frac{\rmd^2}{\rmd{\zeta}^2}
    + 2\,\textrm{sn}_{\pm}^2 - 1
    \right) \varphi_{\pm}^{(n,j)} = m_{(n,j)}^2 \varphi_{\pm}^{(n,j)} \label{eq:NACSL_phi_pm} \,, \\
    \left[
    -\frac{\rmd^2}{\rmd{\zeta}^2}
    -\frac{1}{k^2}(\textrm{dn}_+-\textrm{dn}_-)^2
    -\textrm{cn}_+^2\textrm{cn}_-^2
    +\textrm{sn}_+^2\textrm{sn}_-^2
    \right] \psi_+^{(n,j)} = \mu_{(n,j)}^2 \psi_+^{(n,j)} \label{eq:NACSL_pi_pm} \,.
\end{gather}
Here, we manifestly show the band index $j$ for labeling the eigenvalues $m_{(n,j)}$ and $\mu_{(n,j)}$ with the fixed $n$.

\begin{figure}[tp]
  \begin{center}
   \includegraphics[width=15.0cm]{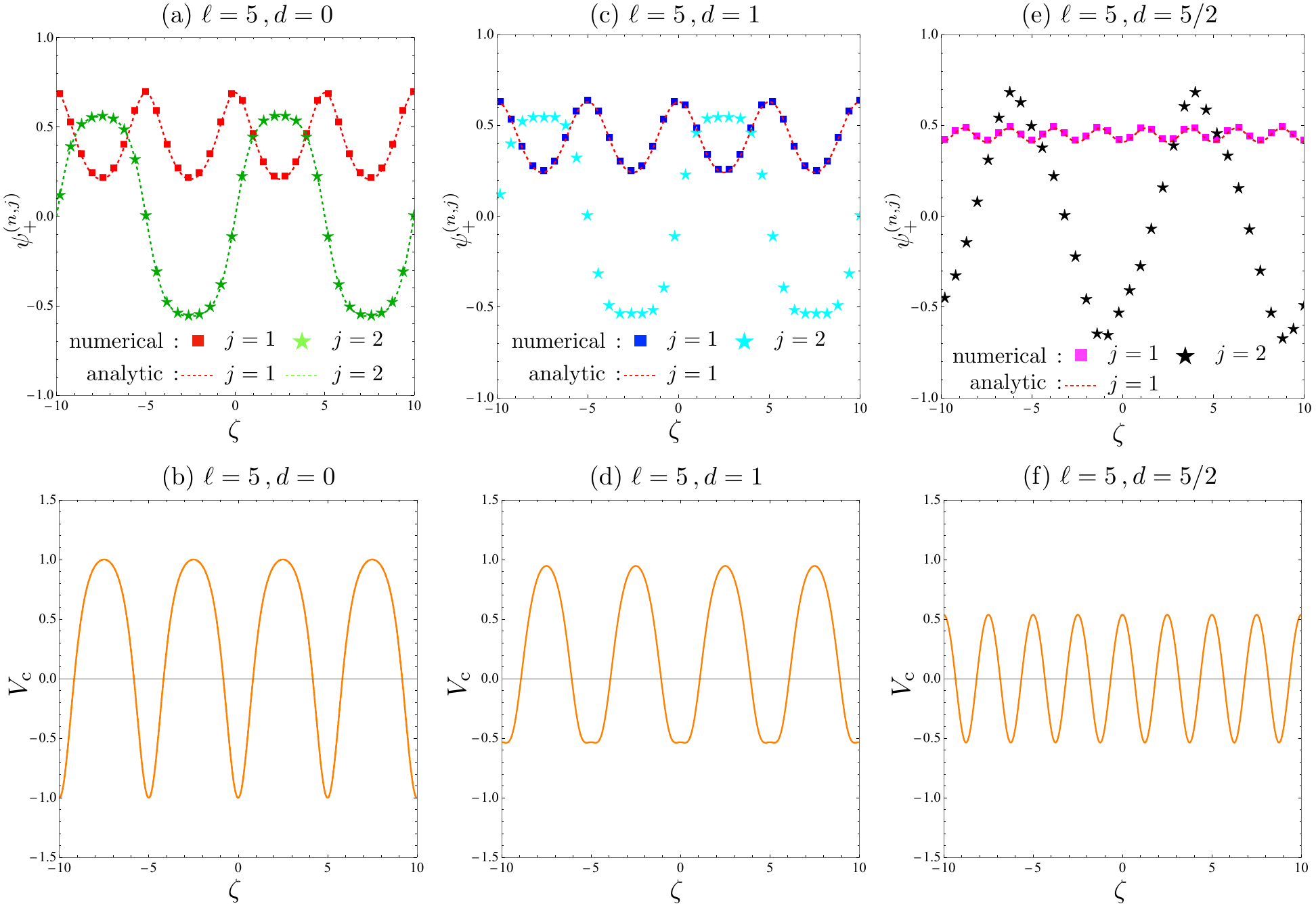}
  \end{center}
  \caption{
  [Noninteracting case: the charged components $\psi_+^{(n,j)}$]
  In the top row,
  $\psi_+^{(0,1)}$ for $d=0,1,5/2$ are shown by the red, blue, and magenta-filled squares, respectively.
  $\psi_+^{(\pi/\ell,2)}$ for $d=0,1,5/2$ are also shown by the green, cyan, and black-filled stars, respectively.
  The analytic solution in eq.~\eqref{eq:lame_massless_psi} of $\psi_+^{(0,1)}$ for all $d=0,1,5/2$ corresponds to the red-dotted lines.
  The analytic solution of \( \psi_+^{(\pi/\ell,2)} \) is only known for the case \( d=0 \), which is shown by
  the green dotted line in (a).
  In the bottom row,
  the potentials $V_{\mathrm{c}}$ for $d=0,1,5/2$ are shown. 
  }
  \label{fig:wavefunction_lame_charged}
\end{figure}

Note that the cases of $d=0$ and $d\neq0$ are qualitatively different.
Since $\phi_+ = \phi_- = \phi_0$ ($\phi_3 = 0$) in the former case, $\varphi_\pm$ and $\psi_+$ satisfy the identical equation. On the other hand,
in the latter case, the neutral components $\varphi_\pm^{(n)}$ satisfy the obviously different equation from those for the charged components $\psi_+^{(n)}$.
Regardless of $d$, we can analytically show that the lowest eigenstates for $n=0$ (corresponding to $\bar p = 0$) and $j=1$ are massless as
\begin{eqnarray}
    &&m_{(0,1)}^2 = 0 \,, \quad
    \varphi_{\pm}^{(0,1)} = \frac{1}{\sqrt{N^{(0,1)}}}{\rm dn} \left(\frac{\zeta\mp d/2}{k},k\right) \,, \label{eq:lame_massless_varphi} \\
    &&\mu_{(0,1)}^2 = 0 \,, \quad
    \psi_+^{(0,1)} = \frac{1}{\sqrt{M^{(0,1)}}}\left[
    {\rm dn} \left(\frac{\zeta - d/2}{k},k\right)
    + {\rm dn} \left(\frac{\zeta + d/2}{k},k\right)
    \right]\,, \label{eq:lame_massless_psi}
\end{eqnarray}
where $N^{(0,1)}$ and $M^{(0,1)}$ are normalization factors defined as
\begin{gather}
    N^{(0,1)} = \int_{0}^{\ell}\rmd \zeta \,
    {\rm dn} \left(\frac{\zeta\mp d/2}{k},k\right)^2
    = 2kE(k) \,, \\
    M^{(0,1)} = \int_{0}^{\ell}\rmd \zeta \,
    \left[
    \left(\frac{\zeta - d/2}{k},k\right)
    + {\rm dn} \left(\frac{\zeta + d/2}{k},k\right)
    \right]^2 \,.
\end{gather}
In figure \ref{fig:wavefunction_lame_neutral} and \ref{fig:wavefunction_lame_charged},
we show these zero modes given in eqs.~(\ref{eq:lame_massless_varphi}) and (\ref{eq:lame_massless_psi}).
They are localized at the potential minima of $V_{\mathrm{n}}$ and $V_{\mathrm{c}}$, respectively.
There are always four massless modes.
Their physical interpretations depend on $d$.
When $d=0$,
the background configuration is the Abelian CSL ($\phi_{1,2,3} = 0$) which is invariant under ${\rm SU(2)_V}$.
One of the four zero modes is the translational zero mode,
whereas the other three are members of the triplet and are the so-called quasi-NG modes 
\cite{Nitta:1998qp,Nitta:2014jta}
behind which there are no spontaneously broken symmetries. 
When $d\neq0$,
the background configuration is the non-Abelian CSL ($\phi_3 \neq 0$).
Then,
$\varphi_{\pm}^{(0)}$ are the phonons propagating independently on the u- and d-CSLs,
corresponding to the spontaneously broken translational symmetries.
Precisely speaking,
the genuine translational zero mode is unique,
and the rest is the quasi-NG mode related to the relative motion of the u- and d-CSLs.
On the other hand,
$\psi_\pm^{(0)}$ is the massless modes associated with the SSB ${\rm SU(2)_V \to U(1)}$,
which is called the isospinon \cite{Eto:2021gyy}.
In both cases of $d=0$ and $d\neq0$,
the total number of zero modes are preserved \cite{Nitta:1998qp}.

\begin{figure}[tp]
  \begin{center}
   \includegraphics[width=11.0cm]{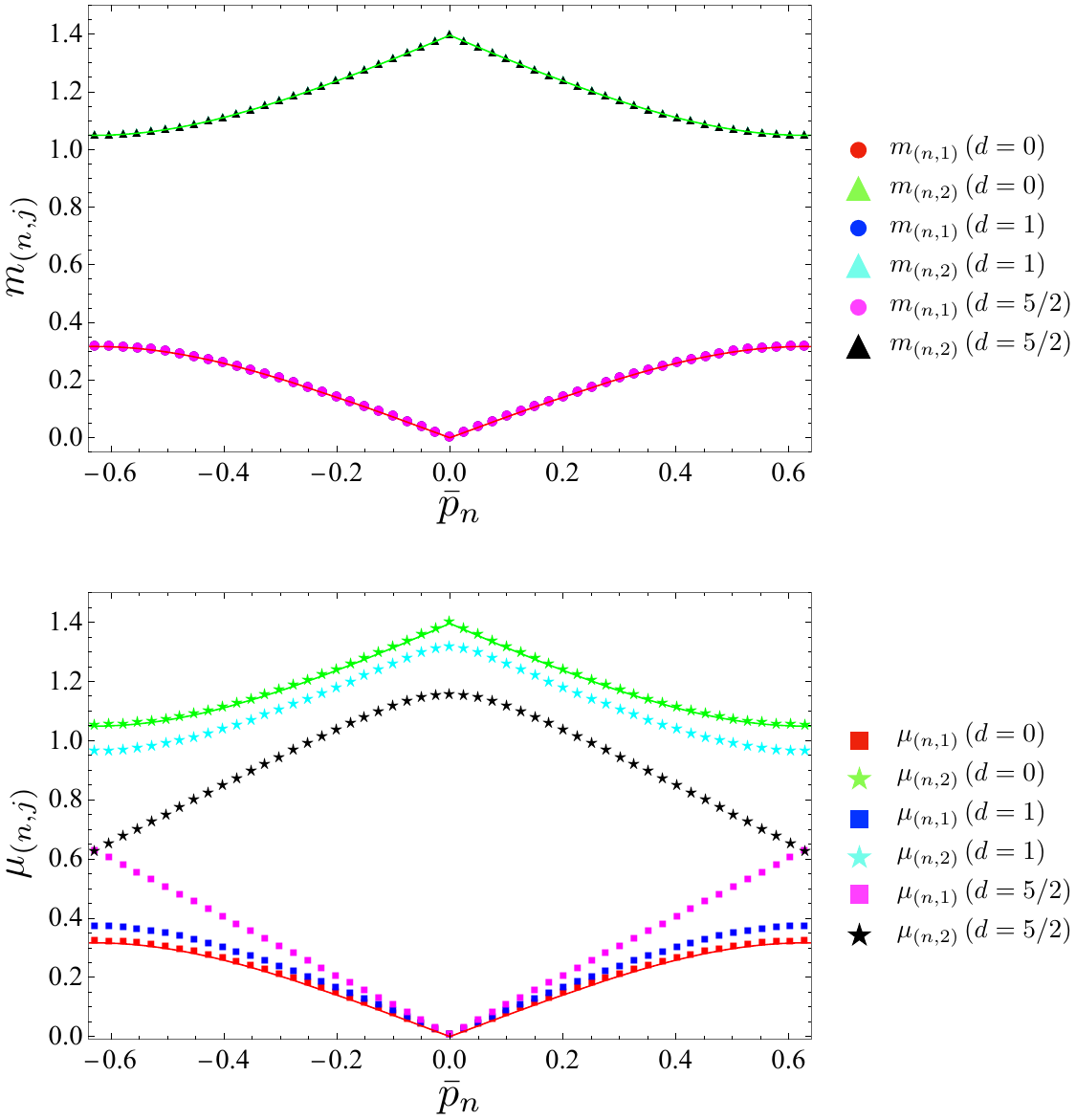}
  \end{center}
  \caption{
  [Noninteracting case]
  Dispersion relations of the phonons $\varphi_{+}^{(n,j)}$ (upper panel) and the isospinons $\psi_+^{(n,j)}$ (bottom panel) for the background CSL with $\ell=5$.
  In the upper panel,
  the $j=1$ ($j=2$) modes at $d=0,1,5/2$ are indicated by red (green), blue (cyan), and magenta (black) circles (triangles), respectively.
  The solid red (green) curves correspond to the analytic solution in eq.~(\ref{eq:analytic_dispersion_valence}) [(\ref{eq:analytic_dispersion_conduction})].
  All three circles (triangles) with different colors completely overlap, reflecting the fact eq.~(\ref{phi_pm_at_ideal}) is independent of $d$.
  In the bottom panel,
  the $j=1$ ($j=2$) modes at $d=0,1,5/2$ are shown by red (green), blue (cyan), and magenta (black) squares (stars), respectively.
  The dispersion relation of the isospinon for $d=0$ coincides with the one for the phonon.
  Consequently, the red squares and the green stars are on the solid red and green curves, respectively.
  }
  \label{fig:dispersion(l1=0,l2=0)}
\end{figure}

Next, let us include a nonzero crystal momentum $\bar p$ and determine the dispersion relation.
The results from the numerical calculations are shown in figure \ref{fig:dispersion(l1=0,l2=0)} for the three different separations $d = 0, 1, 5/2$ for $\ell = 5$ ($d=0$ is the closest case and $d=5/2$ is the farthest case).
In fact, eq.~(\ref{eq:NACSL_phi_pm}) for the neutral components (phonons) can be analytically solved as follows.
It is the so-called Lame equation \cite{Sutherland:1973zz} which can be solved by using the Jacobi theta functions.
For $N=\infty$ (continuum limit), the solution for $j=1$ (called the valence band) is
\begin{eqnarray}
    \bar p = Z(\alpha, k^{\prime}) + \frac{\pi \alpha}{2KK^{\prime}} \,, \qquad
    m_{(\bar p,1)}^2 = \frac{k'{}^2}{k^2} {\rm sn}^2 (\alpha, k') \,,\label{eq:analytic_dispersion_valence}
\end{eqnarray}
where $\alpha$ is an arbitrary real number, and $Z$ is the Jacobi Zeta function.\footnote{
The definition of the Jacobi zeta function is $Z(\alpha,k) = E_k(\alpha) - \frac{E(k)}{K(k)}\alpha$,
with $E_k(\alpha) = \int_{0}^{\alpha}\rmd \alpha'\, \mathrm{dn}^2(\alpha',k)$.
}
Here, we defined $k'=\sqrt{1-k^2}$ and $K' = K(k')$.
In a manner similar to the above,
the solution for \(j=2\) (called the conduction band) is given as 
\begin{gather}
    \bar p = \frac{\pi \alpha}{2KK'} + Z(\alpha, k')
    + {\rm dn}(\alpha, k')\frac{
    {\rm cn}(\alpha, k')
    }{
    {\rm sn}(\alpha, k')
    } \,, \qquad
    m_{(\bar p,2)}^2 = \frac{1}{k^2{\rm sn}^2 (\alpha, k')} \,.\label{eq:analytic_dispersion_conduction}
\end{gather}
The lowest energy solution for $j=2$ is at the edge of the first Brillouin zone $\bar p = \pi/\ell$ and it is analytically given by
\begin{gather}
    \varphi_{\pm}^{(\pi/\ell,2)}
    = \frac{1}{\sqrt{N^{(\pi/\ell,2)}}}\mathrm{sn}
    \left(\frac{\zeta\mp d/2}{k},k\right) \label{eq:lame_massless_varphi_edge} \,, \\
    N^{(\pi/\ell,2)}
    = \int_{0}^{\ell}\rmd \zeta \,
    {\rm sn} \left(\frac{\zeta\mp d/2}{k},k\right)^2
    =\frac{2(K(k)-E(k))}{k} \,.
\end{gather}
We compare the numerical and analytical dispersion relations for the neutral components in the top panel of fig.~\ref{fig:dispersion(l1=0,l2=0)}. They nicely agree with each other.
Note that eq.~(\ref{eq:NACSL_phi_pm}) is independent of $d$, so that the dispersion relations for $d=0,1,5/2$ completely overlap. 
The phonon group velocities are known to be \cite{HuiLi_2000}
\begin{eqnarray}
    c_{\rm ph}^{\rm(u,d)} = \sqrt{1-k^2}\frac{E(k)}{K(k)}\,.
\end{eqnarray}
This vanishes at the single soliton limit with $k=1$ ($B=B_{\rm CSL}$) whereas the rapidity approaches the speed of light as $k\to 0$ ($B\to\infty$) \cite{Brauner:2016pko}.
This behavior agrees with the intuition that the lattice size is $0$ and the CSL is maximally hard at $k=0$ whereas the lattice size is infinite and the CSL is maximally soft at $k=1$.

\begin{figure}[tp]
  \begin{center}
   \includegraphics[width=11.0cm]{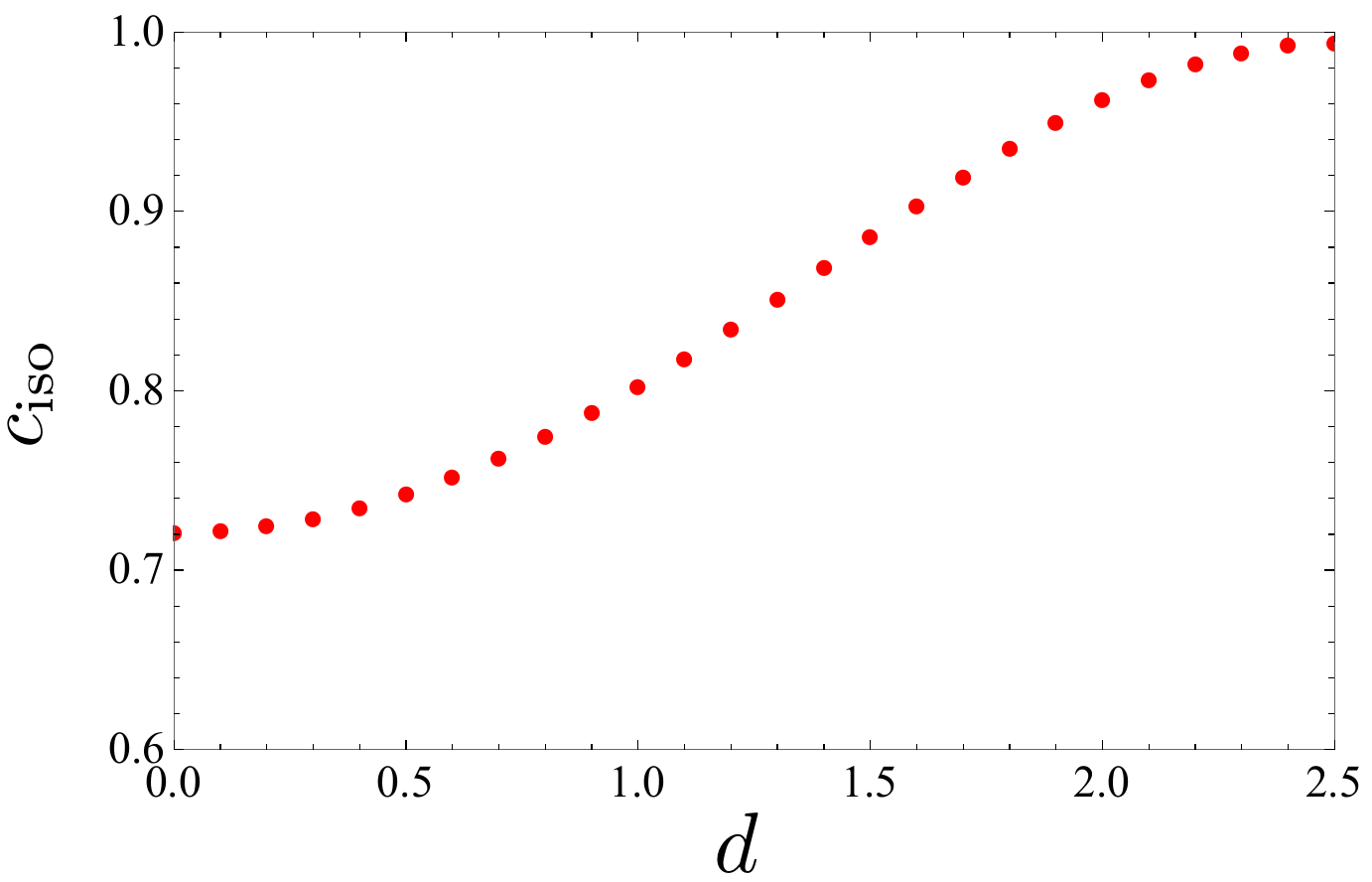}
  \end{center}
  \caption{
  The $d$-dependence of the group velocity of the isospinon at $\bar p =0$ for $\ell=0$, $\epsilon=0$ and $\beta=0$.
  }
  \label{fig:group_velocity}
\end{figure}

In contrast,
the eigen equation (\ref{eq:NACSL_pi_pm}) for the charged components (isospinons) for $0<d\le \ell/2$ cannot be solved analytically, we solve it numerically for the first time to the best of our knowledge. The results are shown in the bottom panel of fig.~\ref{fig:dispersion(l1=0,l2=0)}.
Only when $d=0$, eq.~(\ref{eq:NACSL_pi_pm}) is identical to eq.~(\ref{eq:NACSL_phi_pm}), which can be analytically solved as explained above.
The red squares and green stars in the bottom panel of fig.~\ref{fig:dispersion(l1=0,l2=0)} show the dispersion relations for $j=1$ and $j=2$ of the Abelian CSL ($d=0$).
The dispersion relations of the dimer CSL ($d=1$) 
are shown by the blue squares ($j=1$) and cyan stars ($j=2$).
Similarly, those of the deconfined CSL $(d=5/2)$ correspond to the magenta squares ($j=1$) and black stars ($j=2$).
Interestingly, the gap between $j=1$ and $j=2$ bands at the edge of the first Brillouin zone $\bar p = \pi/\ell$ gradually narrows as increasing the separation $d$, and it closes at $d=\ell/2 = 5/2$ (the black stars and magenta squares).
Furthermore, the dispersion relation of the isospinon at $d=\ell/2$ looks linear and its group velocity seems to reach at the speed of light. This implies that the deconfined CSL ($d=\ell/2$) is the hardest limit for the isospinons. The $d$ dependence of the group velocity of the isospinon at $\bar p=0$ is shown in fig.~\ref{fig:group_velocity}.
Note that the group velocity is usually defined as the momentum derivative of the dispersion relation.
However, since the dispersion relation is currently calculated only for discrete momenta $\bar p_n$,
the group velocity was numerically computed using forward difference:
\begin{gather}
    c_{\mathrm{iso}}
    \equiv \frac{\mu_{(n+1,j)}-\mu_{(n,j)}}{\Delta \bar p_n}
    \,, \qquad
    \Delta \bar p_n \equiv \frac{2\pi(n+1)}{N\ell}
    - \frac{2\pi n}{N\ell} = \frac{2\pi}{N\ell} \,.
\end{gather}

\subsection{Confined phase} \label{subsec:confined}
\begin{figure}[tp]
  \begin{center}
   \includegraphics[width=12.0cm]{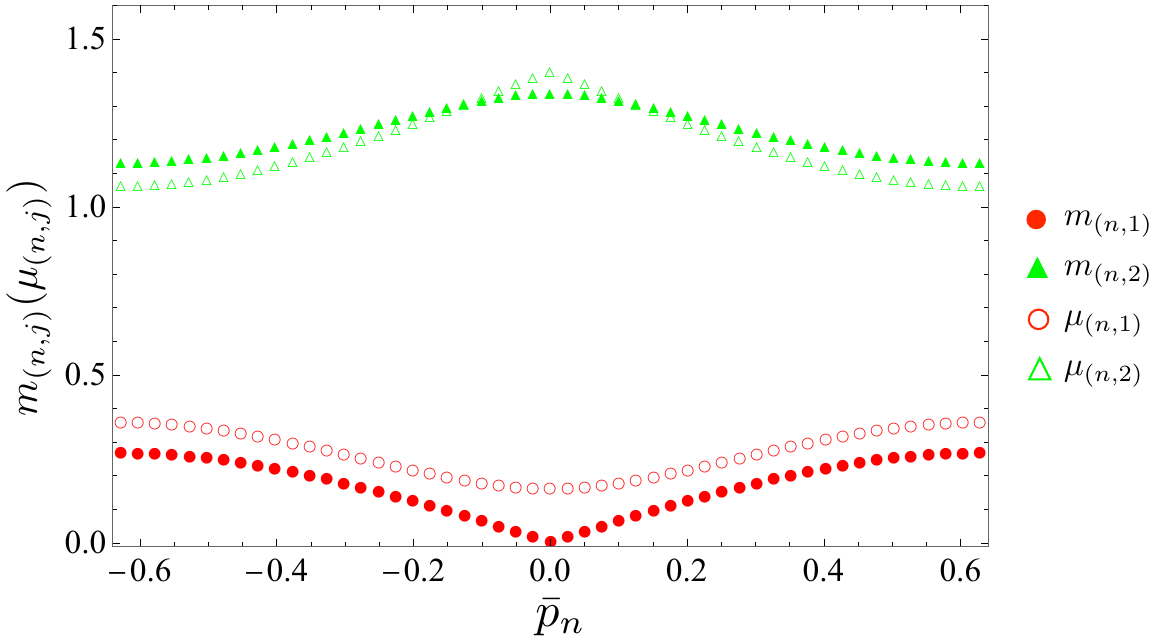}
  \end{center}
  \caption{[Interacting case]
  Dispersion relation of $\varphi_{1,2,3}$ and $\varphi_0$ in eqs.~(\ref{EOM of abelian(pi0)}) and (\ref{EOM of abelian(pi3)})
  for $\epsilon=0$, $\beta=\pi/16$ and $\ell=5$.
  The red-(un)filled circles are the $j=1$ modes of $\varphi_0$ ($\varphi_{1,2,3}$).
  The green-(un)filled triangles are the $j=2$ modes of $\varphi_0$ ($\varphi_{1,2,3}$). 
  }
  \label{fig:dispersion_confined_phase}
\end{figure}

In this phase, the up and down solitons completely adhere, so the separation is $d=0$.
Then, only $\bar \phi_0$ spatially modulates and $\bar \phi_3=0$.
We choose typical parameters for the confined phase: $\epsilon=0\,, \beta=\pi/16\,, \ell=5$.
Here we use $\varphi_{0,1,2,3}$ instead of $\varphi_\pm$ and $\psi_\pm$.
Since the isospin symmetry $SU(2)_{\rm V}$ is unbroken,
$\varphi_{1,2,3}$ form the triplet of $SU(2)_{\rm V}$ and they satisfy the identical equation. 
The EOMs of the fluctuations become
\begin{gather}
    m_{(n,j)}^2 \varphi_0^{(n,j)}
    = \left[
    -\del_{\zeta}^2
    + (4\sin \beta \, \cos(2{\bar \phi}_0)
    + \cos\beta \, \cos{\bar \phi}_0)
    \right]\varphi_0^{(n,j)} \label{EOM of abelian(pi0)} \,, \\
    \mu_{(n,j)}^2 \varphi_{1,2,3}^{(n,j)}
    = \left[
    -(1-\epsilon)\del_{\zeta}^2
    + \cos\beta \, \cos{\bar \phi}_0
    \right]\varphi_{1,2,3}^{(n,j)} \label{EOM of abelian(pi3)} \,.
\end{gather}
We numerically solve these equations,
and their dispersion relations are shown in figure \ref{fig:dispersion_confined_phase}.
Comparing this with the top panel of figure~\ref{fig:dispersion(l1=0,l2=0)},
we find only one phonon in $\varphi_0^{(n,1)}$ which corresponds to the translational zero mode.
The other phonon in $\varphi_3^{(n,1)}$ at $\epsilon=\beta = 0$,
which is the quasi-NG,
is lifted due to the attractive interaction between the up- and down-CSLs.
Because of the isospin symmetry SU$(2)_{\rm V}$, $\varphi_{1,2}$ has the same dispersion relation as $\varphi_3$.
The phonon $\varphi_0^{(n,1)}$ dispersion relation corresponds to the red-filled circle in figure \ref{fig:dispersion_confined_phase}, whereas the red-unfilled one to the lifted quasi-NG mode $\varphi_{1,2,3}^{(n,1)}$.

%

\subsection{Deconfined phase} \label{subsec:deconfined_phase}
\begin{figure}[tp]
  \begin{center}
   \includegraphics[width=10.0cm]{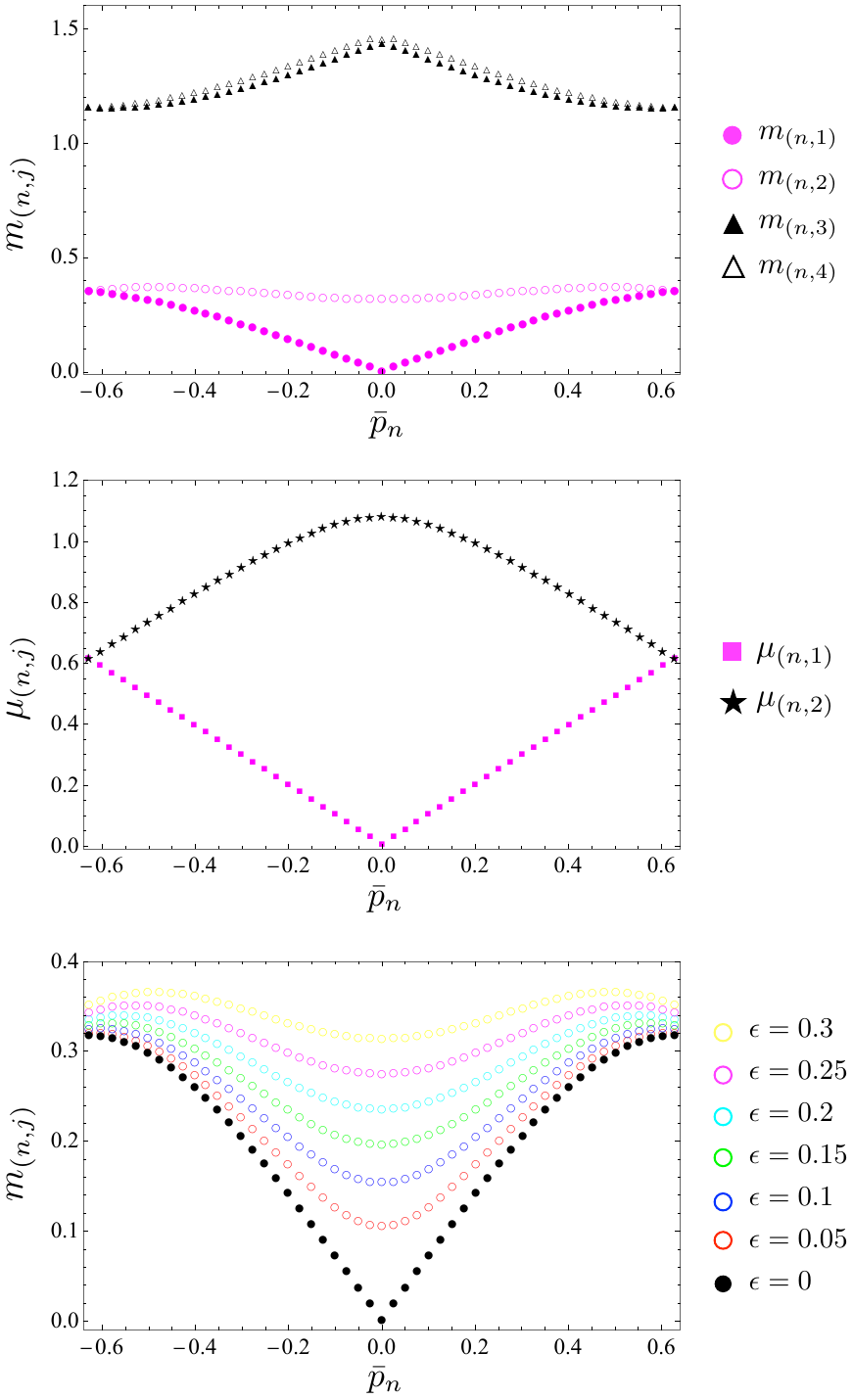}
  \end{center}
  \caption{
  [Interacting case]
  Dispersion relation of the neutral components $\varphi_{\pm}^{(n,j)}$ (upper panel) and charged components $\psi_{\pm}^{(n,j)}$ (middle panel) for $\epsilon=0.3$, $\beta=0$ and $\ell=5$.
  In the upper panel, we show the low-lying four bands $j=1,2,3,4$.
  In the middle panel, we show the low-lying two bands with $j=1,2$ for $\psi_{\pm}^{(n,j)}$ (Each band is doubly degenerate for $+$ and $-$).
  In the bottom panel,
  the $\epsilon$-dependence of $\bar \omega_{(n,2)}$ for $\epsilon=0.05, 0.1, 0.15, 0.2, 0.25, 0.3$ with $\beta=0$ and $\ell=5$ being fixed is shown.
  The black-filled circles coincide with the magenta-filled circles in the top panel, and the yellow-unfilled circles coincide with the magenta-unfilled circles in the top panel.
  }
  \label{fig:dispersion_NACSL_phase}
\end{figure}
The up and down solitons in the deconfined phase are maximally separated; namely, the separation is $d=\ell/2$.
The spatial translational symmetry is spontaneously broken in the deconfined phase as well as the confined one.
In contrast to the two phonons in the noninteractive case of $\epsilon=0$ and $\beta=0$,
there appears to be only one phonon due to the repulsive interaction between the up and down solitons.

We numerically solve eq.~(\ref{eq:EOM_of_u+-}) for $\epsilon \neq 0$ and $\beta=0$ yielding a CSL in the deconfined phase. Note that, when \(\beta = 0\)
the inter-soliton force is generated only by $\epsilon$, and it is repulsive for any positive \(\epsilon\).
The dispersion relations are summarized in figure \ref{fig:dispersion_NACSL_phase}.

The upper panel in figure \ref{fig:dispersion_NACSL_phase} shows that the gapless phonon mode (magenta-filled circle) $\epsilon=0.3$ appears due to the spontaneous translational symmetry breaking.
The gapped mode (magenta-unfilled circle) originates from one of the two gapless phonons present when \(\epsilon=0\) and \(\beta=0\).
Hence, as \(\epsilon\) nears zero,
the dispersion relation for this gapped mode aligns with that of \(\epsilon=0\) and \(\beta=0\) as can be seen in the bottom panel of figure \ref{fig:dispersion_NACSL_phase}.
As for the black-filled and -unfilled triangles,
they tend to be degenerate in the limit of \(\epsilon\) to zero.
As a result, they align with the black triangles depicted in figure \ref{fig:dispersion(l1=0,l2=0)}.

In contrast to the confined phase,
the deconfined phase possesses two NG modes (isospinons) resulting from the spontaneous breaking of the vector SU$(2)_{\rm V}$ symmetry as illustrated in the middle panel of figure \ref{fig:dispersion_NACSL_phase}.
The eigenvalues \(\mu_{n,j}\) exhibit double degeneracy because both \(\psi^{(n,j)}\) and its complex conjugate \(\psi^{(n,j)*}\) adhere to the same equation.
As a result, in the center panel of figure \ref{fig:dispersion_NACSL_phase},
the bands corresponding to \(j=1\) and \(j=2\) are degenerate.
This is similar to the upper panel,
where the first four eigenvalues are illustrated in ascending order.

\subsection{Dimer phase}
\begin{figure}[tp]
  \begin{center}
   \includegraphics[width=15.0cm]{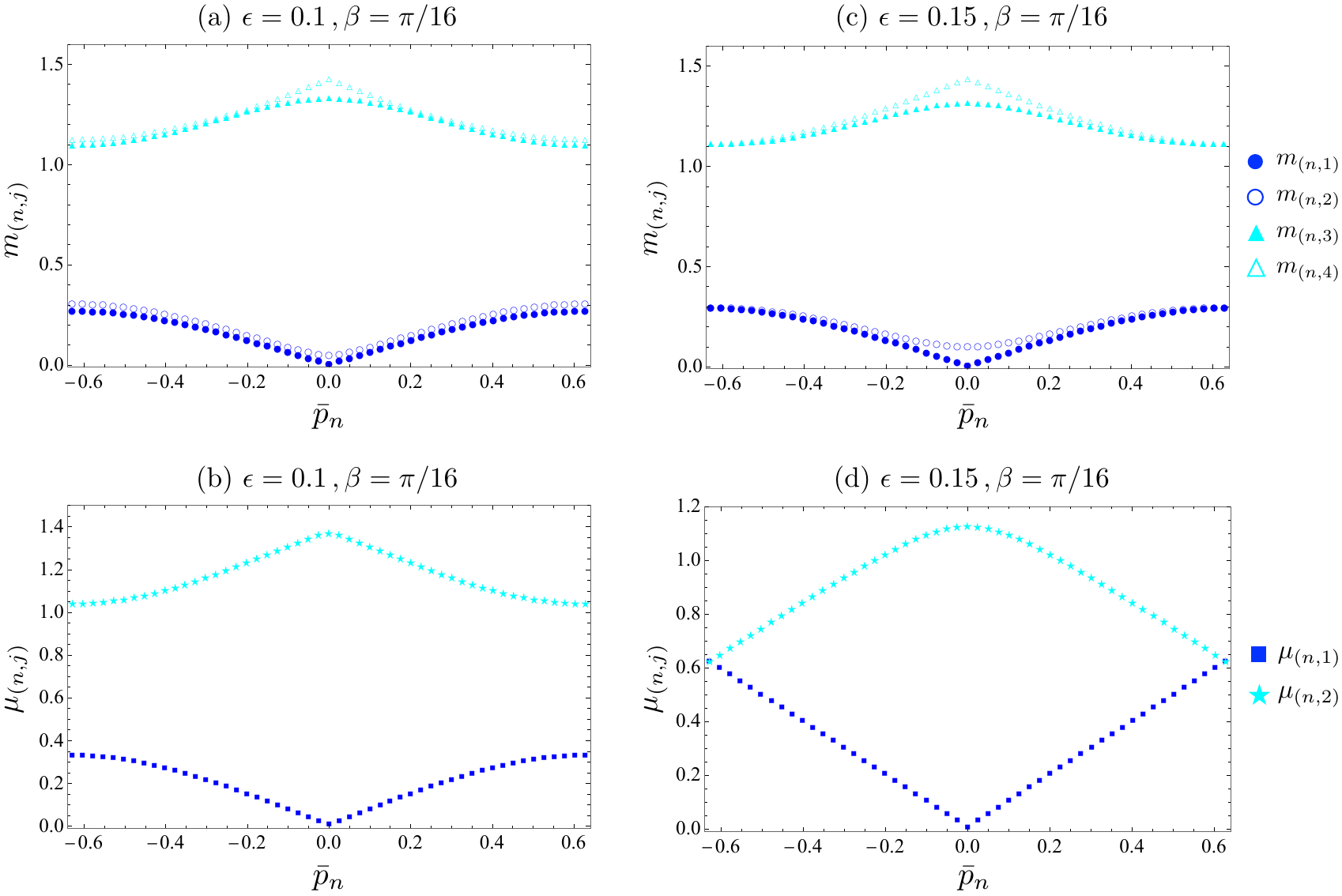}
  \end{center}
  \caption{
  Dispersion relation of $\varphi_{\pm}^{(n,j)}$ and $\psi_{\pm}^{(n,j)}$ in the dimer phase.
  For $\epsilon=0.1$ and $\beta=\pi/16$,
  the dispersion relations of $\varphi_{\pm}^{(n,j)}$ and $\psi_{\pm}^{(n,j)}$ are (a) and (b), respectively.
  For $\epsilon=0.15$ and $\beta=\pi/16$,
  the dispersion relations of $\varphi_{\pm}^{(n,j)}$ and $\psi_{\pm}^{(n,j)}$ are (c) and (d), respectively.
  }
  \label{fig:dispersion_dimer_phase}
\end{figure}

As discussed in section \ref{sec:NACSL},
the deconfined and dimer phases share the same pattern of continuous symmetry breaking,
with their only distinction lying in the discrete symmetry.
Consequently, the emergence of a single phonon and two isospinons in the dimer phase is naturally expected.

We again numerically solve equations \eqref{eq:EOM_of_u+-} and \eqref{eq:g+} for \(\epsilon=0.1\) and \(\epsilon=0.15\). The lattice constant is again fixed to be
\(\ell=5\), and we now turn on \(\beta=\pi/16\).
We numerically verified the emergence of a single phonon,
resulting from the spontaneous breaking of spatial translational symmetry,
which is depicted by the blue-filled circles in (a) and (c) of figure \ref{fig:dispersion_dimer_phase}.
We also confirmed the presence of two isospinons,
as indicated by the blue-filled squares in (b) and (d) of figure \ref{fig:dispersion_dimer_phase}.
These are comparable to those observed in the top and middle panels of figure \ref{fig:dispersion_NACSL_phase} for the deconfined phase.
A notable distinction, however,
is evident at the edge of the Brillouin zone (\(\bar p_n = \pm \pi/\ell\)). 
The two bands with \(j=1\) and \(j=2\) close there in the deconfined case whereas there is a finite gap for the dimer case.
Comparing (b) and (d) in figure \ref{fig:dispersion_dimer_phase}, it clearly shows that the band gap narrows as $\epsilon$ increases.
This can be understood as follows.
When we increase \(\epsilon\), the repulsion between the up and down solitons becomes stronger, so that the dimer becomes longer. 
Consequently, at some critical $\epsilon$,
the transition from the dimer phase to the deconfined phase occurs.
We have seen a similar phenomenon for \(\epsilon=0\) and \(\beta=0\),
as demonstrated in the bottom panel of figure \ref{fig:dispersion(l1=0,l2=0)},
where the band gap narrows as \(d\) increases.

\section{Summary and discussion}\label{sec:summary}
We have numerically studied excitations and their dispersion relations around Abelian and non-Abelian CSLs in the confined, deconfined, and dimer phases of rapidly rotating QCD matter.
We have found three gapless type-A NG modes with linear dispersion relations: two isospinons (the ${\mathbb C}P^1$ modes) and a phonon corresponding to the spontaneously broken vector SU$(2)_{\rm V}$ and translational symmetries in non-Abelian CSLs (deconfined and dimer phases).
We also have found only one gapless NG mode (a phonon) in the confined phase since the SU$(2)_{\rm V}$ symmetry is recovered.
We have obtained gapped modes (first excited modes) as well in these phases. 
The number of gapless NG modes is different between Abelian ($\eta$) and non-Abelian CSLs while those cannot distinguish the two non-Abelian CSLs,
the deconfined and dimer phases since they are distinct only in terms of the discrete symmetry. 
We also have found in the deconfined phase that the dispersion relation of the isospinons becomes of the Dirac type, {\it i.~e.~} linear even at large momentum.

Here, we address some discussions.
The first is about the types of NG modes.
In the non-Abelian CSLs,
the ${\mathbb C}P^1$ modes of neighboring solitons repel 
and are antialigned 
at least in the parameters that we studied. 
Thus, they behave as an anti-ferromagnetic Heisenberg model.
It is an open question whether there is the case that 
the ${\mathbb C}P^1$ modes of 
neighboring solitons attract each other and tend to be aligned, 
constructing a soliton lattice behaving 
as a ferromagnet. 
This question is related to the type of NG modes;
In such a case, the ${\mathbb C}P^1$ modes would be of type-B 
with a quadratic dispersion relation.
In this regard, 
type-B NG modes often appear in finite density systems which are nonrelativistic.
For instance, in the presence of a single domain wall 
in the nonrelativistic ${\mathbb C}P^1$ model, 
the translational zero mode is coupled 
with an internal mode to become a type-B NG mode 
\cite{Kobayashi:2014xua,
Kobayashi:2014eqa}.\footnote{
It is also known that in the presence of a soliton, 
translational modes of the soliton 
sometimes can have fractional dispersion relations 
\cite{Watanabe:2014zza,Takahashi:2015caa}. This seems to be the case that zero modes are nonnormalizable 
and is not applied to CSLs.
}
Whether type-B NG modes can appear in a CSL is 
one of the directions to explore. 

In this paper, we have not considered a coupling to the 
electromagnetism. When the electromagnetic coupling is taken into account 
which is more relevant to the case of neutron stars and heavy-ion collisions,
a photon has quadratic dispersion relation (as a type-B NG mode) 
\cite{Yamamoto:2015maz}.
Such nonrelativistic photon 
should be discussed in a 
non-Abelian CSL background.

The extension to the three-flavor case 
(the up, down, and strange quarks) 
is one of the important directions. 
In this case, 
the $\eta'$ meson has the anomalous coupling with rotation,
and the Abelian $\eta'$-CSL was studied before 
\cite{Nishimura:2020odq}. 
In some parameter regions, 
a single $\eta'$-soliton is split into
three non-Abelian solitons (the deconfined phase).
In such a case, the vector symmetry 
SU$(3)_{\rm V}$ is spontaneously broken 
to an SU$(2) \times U(1)$ subgroup
in the vicinity of each non-Abelian soliton 
so that 
there appear ${\mathbb C}P^2 \simeq 
{\rm SU}(3)_{\rm V}/[{\rm SU}(2) \times {\rm U}(1)]$ zero modes 
localized around each soliton. 
If the ${\mathbb C}P^2$ modes of neighboring solitons repel as the case of two flavors 
in the deconfined phase,
then the lattice can be regarded as an antiferromagnetic 
SU$(3)$ Heisenberg model. 
A continuum limit for extremely rapid rotation would result in an SU$(3)/$U$(1)^2$ flag manifold nonlinear sigma model.
Thus, we expect six type-A NG modes and one phonon in the three-flavor case. 

CSL phases (also called spiral phases) are also present in chiral magnets 
accompanied by the Dzyaloshinskii-Moriya (DM) interaction
\cite{Dzyaloshinskii,Moriya:1960zz}, 
and 
have been paid great attention 
\cite{togawa2012chiral,togawa2016symmetry,KISHINE20151,
PhysRevB.97.184303,PhysRevB.65.064433,Ross:2020orc,Amari:2023gqv}. 
Let us discuss the similarities and differences between 
the CSL phases in rotating QCD matter studied in this paper and those in chiral magnets. 
In the former, 
we have found isospinons (${\mathbb C}P^2$ modes) 
besides a phonon, 
and these modes are decoupled from each other. 
On the other hand, in the case of chiral magnets, 
each domain wall or soliton carries a U(1) modulus in contrast to ${\mathbb C}P^2$ moduli 
for the CSLs in rotating QCD matter. 
Thus, a soliton lattice is the XY model rather than a Heisenberg model in the CSLs in rotating QCD matter. 
If the potential is an easy-axis potential,
the U(1) moduli of the domain walls are alternate in the ground state, and so the XY model is antiferromagnetic. 
In the CSL, there should be a magnon corresponding 
to the U(1) symmetry breaking
beside a phonon corresponding to translational symmetry breaking. 
However, the U(1) magnon and phonon are coupled nontrivially due to the DM interaction 
\cite{Ross:2022vsa,Amari:2023bmx}, 
unlike the case of the CSL in the QCD matter 
in which case 
the ${\mathbb C}P^2$ isospinons and phonon are decoupled. Therefore, their type of NG bosons and dispersion relations are quite nontrivial  problems to be explored in the future.

In a different context,
a lattice of non-Abelian vortices accompanied with ${\mathbb C}P^{N-1}$ modes was studied \cite{Kobayashi:2013axa} instead of the non-Abelian sine-Gordon solitons considered in this paper.
As in the case of non-Abelian CSLs,
the interaction between neighboring vortices intermediates the propagation of the ${\mathbb C}P^{N-1}$ modes into two transverse directions of the lattice in addition to a phonon called the Tkachenko mode \cite{Tkachenko1,Tkachenko2,Tkachenko3,Du:2022xys}. 
In ref.~\cite{Kobayashi:2013axa},
a continuum limit of the lattice 
was studied at a large distance to yield 
an anisotropic ${\mathbb C}P^{N-1}$ model.
Such a continuum limit should be considered for
a non-Abelian CSL in QCD matter as well 
to yield an anisotropic ${\mathbb C}P^{N-1}$ model.

\begin{acknowledgments}
This work is supported in part by 
 JSPS KAKENHI [Grants 
No. JP21H01084 (KN), 
 and No. JP22H01221 (ME and MN)] and the WPI program ``Sustainability with Knotted Chiral Meta Matter (SKCM$^2$)'' at Hiroshima University (KN and MN).
\end{acknowledgments}


\bibliographystyle{jhep}
\bibliography{reference.bib}


\end{document}